\documentclass[12pt]{article}

\usepackage[nokeyprefix]{refstyle}
\usepackage{varioref}
\usepackage{xr-hyper}
\usepackage[colorlinks,citecolor=blue,urlcolor=blue]{hyperref}

\usepackage[margin=1in]{geometry}
\RequirePackage[OT1]{fontenc}
\RequirePackage{amsthm,amsmath,geometry,amsfonts,mathtools}
\RequirePackage[round]{natbib}
\usepackage{longtable}
\usepackage{float, graphicx, enumitem, caption}

\usepackage[latin2]{inputenc}

\usepackage{longtable}
\usepackage{array}
\usepackage{multirow}
\usepackage{amssymb}

\usepackage{soul}
\usepackage{xcolor}

\newcommand{\bD}{\boldsymbol{D}}
\newcommand{\bV}{\boldsymbol{V}}
\newcommand{\bv}{\boldsymbol{v}}
\newcommand{\bZ}{\boldsymbol{Z}}
\newcommand{\bAone}{\boldsymbol{A}_1}
\newcommand{\bAtwo}{\boldsymbol{A}_2}
\newcommand{\bAthree}{\boldsymbol{A}_3}
\newcommand{\bB}{\boldsymbol{B}}
\newcommand{\bX}{\boldsymbol{X}}
\newcommand{\bx}{\boldsymbol{x}}
\newcommand{\bPsi}{\boldsymbol{\Psi}}

\newcommand{\independent}{\perp\!\!\!\perp}
\newtheorem{theorem}{Theorem}

\newtheorem{assumption}{Assumption}

\usepackage{apptools}
\AtAppendix{\counterwithin{assumption}{section}}

\allowdisplaybreaks

\title{Improved inference for doubly robust estimators of heterogeneous treatment effects}
\author{Heejun Shin and Joseph Antonelli \footnote{Heejun Shin, PhD Student, Department of Statistics, University of Florida (Email: hshin1@ufl.edu). Joseph Antonelli, Assistant Professor, Department of Statistics, University of Florida (Email: jantonelli@ufl.edu)}}
\date{}

\begin{document}

\maketitle

\begin{abstract}
    We propose a doubly robust approach to characterizing treatment effect heterogeneity in observational studies. We develop a frequentist inferential procedure that utilizes posterior distributions for both the propensity score and outcome regression models to provide valid inference on the conditional average treatment effect even when high-dimensional or nonparametric models are used. We show that our approach leads to conservative inference in finite samples or under model misspecification, and provides a consistent variance estimator when both models are correctly specified. In simulations, we illustrate the utility of these results in difficult settings such as high-dimensional covariate spaces or highly flexible models for the propensity score and outcome regression. Lastly, we analyze environmental exposure data from NHANES to identify how the effects of these exposures vary by subject-level characteristics.
\end{abstract}

\noindent%
\vfill

\section{Introduction}

Understanding how the effect of a treatment varies across subgroups of the population is a common scientific goal in a wide array of fields. This variation in the treatment effect, typically referred to as treatment effect heterogeneity, is important for learning which individuals are most likely to benefit from treatment.  Analyzing treatment effect heterogeneity is, however, a challenging problem, especially when the dimension of the covariates is large or nonparametric models are employed. One difficulty is the ability to perform inference in such settings as the bootstrap may not be valid, and approaches relying on asymptotic approximations may not perform well, particularly when the sample size is small. To address these problems, we combine posterior distributions for nuisance parameters with doubly robust estimators to provide an estimate of the conditional average treatment effect (CATE) that is consistent if either the outcome or treatment models are correctly specified. We also introduce a novel variance estimation procedure which (\romannumeral1) is consistent when both nuisance models are correctly specified with sufficiently fast contraction rates, and (\romannumeral2) is conservative in finite samples, under model misspecification, or when nuisance parameters are estimated with slow contraction rates.

There has been a significant increase in attention to treatment effect heterogeneity in recent years. A number of approaches fit into a category referred to as meta-learners, which permit the usage of existing estimators in the machine learning or high-dimensional literature when estimating the CATE. Examples of such approaches can be found in \cite{Kunzel2019} and \cite{Nie2021}. 
Other approaches have specifically tailored algorithms towards treatment effect heterogeneity. The causal forest algorithm developed in \cite{Wager2018} uses regression trees and the sample splitting approach of \cite{Athey2016} to provide nonparametric estimation and honest inference of heterogeneous treatment effects. The Bayesian causal forest approach of \cite{Hahn2020} extends the initial work of \cite{Hill2011} that uses Bayesian additive regression trees (BART, \cite{Chipman2012}) to estimate heterogeneous treatment effects. They separate the outcome regression function into two components, one of which corresponds to the CATE, which allows them to focus one BART prior distribution on treatment effect heterogeneity and shrink effects towards an overall homogeneous effect. This approach, or variants of it, have also been shown to work well in recent causal inference data analysis competitions \citep{Dorie2019}. 

Recent work has looked to extend these approaches to high-dimensional situations \citep{Powers2018}. Estimation of average treatment effects in high-dimensional scenarios has garnered substantial interest \citep{belloni2014high,Farrell2015, Chernozhukov2018, antonelli2018doubly, antonelli2019high, antonelli2020averaging, ning2020robust, tan2020regularized}. When estimating average causal effects, adjusting for a high-dimensional set of covariates is a nuisance parameter, and the target of interest is a low-dimensional quantity. The problem is more difficult when estimating the CATE, because the parameter of interest can itself be high-dimensional. One approach is to allow the CATE to depend on a pre-specified subset of the covariates. \cite{Abrevaya2015} suggest using an inverse probability weighted method and integrating out the remaining covariates. However, this estimator is unstable, especially when the propensity score is misspecified, which led to the development of doubly robust estimators in recent papers \citep{Fan2020,kennedy2020optimal,knaus2020double,Lee2017,Semenova2021}. There also has been work on finding effect modifiers using machine learning methods \citep{Athey2016}, especially in the causal rule framework \citep{lee2020causal,Lee2021,Wang2022}.

At the core of much of the work in nonparametric or high-dimensional causal inference are doubly robust estimators \citep{scharfstein1999adjusting, bang2005doubly}. These estimators have been popular for many years due to the namesake property that only one of the propensity score or outcome regression models need be correctly specified to obtain consistent inference. In nonparametric or high-dimensional settings, they have an added advantage that parametric convergence rates can be obtained even when each of the propensity score or outcome regression models converge at slower rates. While this is advantageous for point estimation, inference is more challenging, as confidence intervals commonly rely on both models being correctly specified. Although recent work has aimed to alleviate this \citep{van2014targeted,benkeser2017doubly,dukes2020doubly, tan2020model, avagyan2021high,dukes2021doubly}, nearly all of this work has focused on estimating average treatment effects, while the focus of this paper is on conditional average treatment effects. 

In this paper, we develop a novel frequentist approach that combines Bayesian modeling of the propensity score and outcome regression models with meta learners for estimating heterogeneous treatment effects. This approach allows us to handle high-dimensional confounder spaces or utilize highly flexible nonparametric Bayesian models. This extends the literature on doubly robust estimation and doubly robust inference as we are able to provide improved inference on conditional average treatment effects even in finite samples or model misspecification. We see empirically in Section \ref{sec:Simulations} that this leads to substantial gains in confidence interval performance. In conjuction with flexible nonparametric Bayesian models that ensure small biases of resulting treatment effect estimates, this leads to a procedure with strong estimation and inferential properties. 

\section{Estimands and Identifying Assumptions}

Our goal will be to estimate the effect of a treatment $T$ on an outcome $Y$, and to understand whether the treatment effect varies by observed characteristics. We observe a $p-$dimensional vector of covariates denoted by $\bX$ that are used to adjust for confounding bias. Throughout, we will allow $p$ to be large and potentially growing with the sample size, denoted by $n$. We are interested in learning the extent to which the treatment effect varies by covariates $\bV$, where $\bV$ is a $q-$dimensional set of covariates, and $q$ is assumed to be relatively small and not growing with $n$. Typically the covariates in $\bV$ will be a subset of the covariates in $\bX$, though the remaining ideas hold even if they are not a subset of the covariates in $\bX$. We assume that we observe $n$ independent and identically distributed random variables denoted by $\bD_i = [Y_i, T_i, \bX_i, \bV_i]$ for $i=1,\dots, n$. We are interested in the conditional average treatment effect (CATE) denoted by
$$E(Y(1) - Y(0) | \bV = \bv) = \tau(\bv),$$
where $Y(t)$ is the potential outcome we would observe under treatment level $t$. In order to identify these treatment effects from the observed data we rely on certain assumptions. First, we assume the stable unit treatment value assumption (SUTVA, \cite{little2000causal}). This assumption states that the treatment of one unit does not affect the outcomes of other units and that the treatment is well defined in the sense that $Y_i = Y_i(T_i). $ We also assume positivity and unconfoundedness, which are defined as \\
\\
\textit{Positivity:} $0 < P(T = 1 \vert \bX = \bx) < 1$ for all $\bx$.  \\
\textit{Unconfoundedness:} $Y(t) \independent T | \bX$ for $t=0,1$. \\
\\
Here $P(T = 1 \vert \bX = \bx)$ denotes the propensity score, and positivity states that all subjects have a positive probability of receiving either treatment level. Unconfoundedness effectively states that there are no unmeasured common causes of the treatment and outcome. While there are multiple ways to identify the CATE from the observed data under these assumptions, we will use an approach that relies on constructing a pseudo-outcome defined by
$$Z_i \equiv Z(\bD_i, p_{ti}, m_{ti}) = \frac{1(T_i = 1)}{p_{1i}} (Y_i - m_{1i}) + m_{1i} - \frac{1(T_i = 0)}{p_{0i}} (Y_i - m_{0i}) - m_{0i},$$
where $p_{ti} = P(T = t | \bX = \bX_i)$ and $m_{ti} = E(Y | T = t, \bX = \bX_i)$. Note that for brevity we will typically refer to this quantity simply as $Z_i$, though it is implied throughout that it is dependent on the data $\bD$ and unknown model parameters $\bPsi$.  This pseudo-outcome is doubly robust in the sense that when either $P(T=t | \bX = \bx)$ or $E(Y | T=t, \bX=\bx)$ are correctly specified, the following result holds:
$$E(Z \vert \bV=\bv) = \tau(\bv).$$
Identification through this pseudo-outcome points to a two-step strategy towards estimating causal effects. At the first step, the propensity score and outcome regression models are estimated and the pseudo-outcome is constructed. At the second stage, the pseudo-outcome is regressed against the covariates $\bV$ to estimate $\tau(\cdot)$. We detail our strategy for these two steps, and how we account for all sources of uncertainty in the following section. 

\section{Methodology}

Throughout, we will be working under the assumption that both the treatment and outcome models are fit using Bayesian approaches and therefore we have posterior distributions of both $p_{ti}$ and $m_{ti}$ for $i = 1, \dots, n$ and $t \in \{0,1\}$. We denote all unknown parameters of these two models by $\bPsi$. Our framework is intended to work in the more difficult scenarios when $\bPsi$ is high-dimensional, either because the number of covariates is large relative to the sample size (e.g. $p > n$) or because very flexible, nonparametric approaches have been used to estimate the treatment model, outcome model, or both. These could include nonparametric approaches such as the BART prior, Gaussian process regression, and dirichlet process mixtures, or high-dimensional models such as those based on spike-and-slab prior distributions. Inference in this situation is complicated by three key factors: 1) the estimated treatment effect is not solely a function of the unknown parameters and therefore we cannot simply use the posterior distribution for inference, 2) the bootstrap frequently does not apply when using high-dimensional models (\cite{el2018can}; see Appendix \ref{sec: Bootstrap} for an empirical illustration), and 3) as we see in Section \ref{sec:Simulations}, estimators relying on asymptotic approximations to inference may perform poorly in difficult situations such as finite sample sizes, model misspecification, or high-dimensionality. Our goal is to construct an approach that can provide valid inference despite the presence of these three complicating factors.

\subsection{Estimation and Inferential Strategy}\label{sec:Estimation and inferential strategy}

Once the posterior distribution of the treatment and outcome model parameters are obtained, which we denote by $\bPsi | \bD$, we must construct an estimator of $\tau(\bv)$. Throughout we assume that $\tau(\boldsymbol{v}) = \boldsymbol{v \alpha}^*$ for some $\boldsymbol{\alpha}^*$. This implies the true CATE is a linear function of $\boldsymbol{v}$, though this can include nonlinear functions of the covariates. We can define an estimator of this quantity as follows:
\begin{align}
    \Delta(\bPsi, \bD) = \bv (\bV^T \bV)^{-1} \bV^T \bZ, \label{eqn:DeltaDef}
\end{align}
where $\bZ = [Z_1, \dots, Z_n]^T$ and $\bV$ is the $n \times q$ matrix of treatment effect modifiers. Note that this is a function of both $\bPsi$ and $\bD$ since $Z_i$ is defined to be a function of unknown parameters $\bPsi$ and $\boldsymbol{D}_i$

There are two natural estimators of $\tau(\bv)$ once the posterior distribution is obtained. The more common approach in the causal inference literature is to construct an estimate of both $p_{ti}$ and $m_{ti}$. In our setting, we could use the posterior mean by setting $\widehat{p}_{ti} = E_{\bPsi | \bD} [p_{ti}]$ and $\widehat{m}_{ti} = E_{\bPsi | \bD} [m_{ti}]$. Then, we can plug-in these values to define the pseudo-outcomes, $\widehat{Z}_i = Z(\bD_i, \widehat{p}_{ti}, \widehat{m}_{ti})$ for $i=1, \dots, n$, which we can regress against the effect modifying covariates $\bV$. Using the notation above, this can be defined as $\widehat{\tau}(\bv) = \Delta(\widehat{\bPsi}, \bD)$ with $\widehat{\bPsi} = E_{\bPsi | \bD}[\bPsi]$. While this strategy is reasonable and potentially leads to good performance of the resulting estimates of $\tau(\bv)$, inference is more challenging. Typically, either the bootstrap would be used to account for uncertainty in both stages of the estimator, or the estimator's asymptotic distribution is obtained from which inference can proceed. As discussed above, however, these approaches either may not be valid or may not perform well in finite samples with complex models for the propensity score and outcome regression models. One may think that the posterior distribution can be used to help with regards to inference, but it is not clear how the posterior distribution can be used to properly account for uncertainty in this estimator. For these reasons, we take a second approach to estimating the CATE, which is to construct an estimator as
$$\widehat{\tau}(\bv) = E_{\bPsi | \bD}[\Delta(\bPsi, \bD)],$$
which is the posterior mean of the $\Delta(\bPsi, \bD)$ function. Intuitively, for every posterior draw of the propensity score and outcome regression models, a new pseudo-outcome is constructed, and this outcome is regressed against $\bV$. This is done for every posterior draw, and the mean of these values is our estimator. We approximate this posterior mean using $B$ posterior draws as follows:
\begin{align}
\label{eqn:PointEstimator}
   \frac{1}{B} \sum_{b=1}^B \Delta(\bPsi^{(b)}, \bD) = \frac{1}{B} \sum_{b=1}^B \bv (\bV^T \bV)^{-1} \bV^T \bZ^{(b)}, 
\end{align}
where $Z^{(b)}$ is the $b^{th}$ posterior draw of the pseudo-outcome and is defined as
$$\frac{1(T_i = 1)}{p_{1i}^{(b)}} (Y_i - m_{1i}^{(b)}) + m_{1i}^{(b)} - \frac{1(T_i = 0)}{p_{0i}^{(b)}} (Y_i - m_{0i}^{(b)}) - m_{0i}^{(b)}.$$

Now that we have defined our estimator, we can describe our strategy to estimating the variance of this estimator in a way that accounts for all sources of uncertainty. The true variance of interest is the variance of the sampling distribution of this estimator, and is defined by $\text{Var}_{\bD} E_{\bPsi | \bD}[\Delta(\bPsi, \bD)]$. There are two main sources of variability in this estimator: 1) the uncertainty in parameter estimation for the propensity score and outcome regression models, and 2) sampling variability in $\bD_i$ that is present even if we knew the true outcome and propensity score models. We extend the results seen in \cite{antonelli2022causal} in order to construct an estimator of the variance that separately targets these two sources of variability. We will provide intuition for this variance estimator, and in the following section will show that it is a consistent estimator of the variance that will be conservative in finite samples or under model misspecification.

Before defining our variance estimator, we must introduce additional notation. We will let $\bD^{(m)}$ be a resampled version of our original data $\bD$, where resampling is done with replacement as in the nonparametric bootstrap \citep{efron1994introduction}. Our variance estimator can then be defined as
\begin{align}
    \text{Var}_{\bD^{(m)}} \{ E_{\bPsi | \bD}[\Delta(\bPsi, \bD^{(m)})] \} + \text{Var}_{\bPsi | \bD}[\Delta(\bPsi, \bD)]. \label{eqn:VarEstimator} 
\end{align}
The first of these two terms resembles the true variance, except the outer variance is no longer with respect to $\bD$, but is now with respect to $\bD^{(m)}$. This is a crucial difference, however, as the first term does not account for variability due to parameter estimation. The inner expectation of the first term is with respect to the posterior distribution of $\bPsi$ given the observed data $\bD$, and not the resampled data $\bD^{(m)}$. This means that this variance term does not account for variability that is caused by the fact that different data sets would lead to different posterior distributions. Ignoring this source of variability will likely lead to anti-conservative inference as our estimated variance will be smaller than the true variance of our estimator. To counter this issue, the second term is added, which is the variability of the estimator due to parameter uncertainty. It makes sense to add posterior variability to the first term, which was ignoring uncertainty from parameter estimation, however, it is not clear that the summation of these two terms leads to a valid variance estimator. In Section \ref{sec:variance estimation}, we detail how this variance estimator 1) leads to conservative estimates of the variance in general, and 2) is consistent when both the treatment and outcome models are correctly specified and the posterior distributions of the propensity score and the conditional mean outcome contract at sufficiently fast rates.

\subsection{Properties of the CATE estimator}

First, we detail the properties of our estimator when either the propensity score or outcome regression models is correctly specified. Note that at times, we utilize sample splitting for theoretical proofs where half of the sample is used for estimating the unknown parameters $\boldsymbol{\Psi}$, and the remaining half is used in the second stage of our estimator of the CATE. This simplifies certain calculations substantially, though we expect the same results to hold in the absence of sample splitting. For this reason, we utilize both a cross-fitted estimator in the spirit of \cite{Chernozhukov2018}, as well as one that does not use sample splitting. Empirically we find that the theoretical results hold for both estimators, but the finite sample performance of the estimator without sample splitting is better. Results comparing these two estimators can be found in Appendix \ref{sec:with and without sample splitting}.

Let $\boldsymbol{p}_t = (p_{t1}, \dots, p_{tn})$, $\boldsymbol{m}_t = (m_{t1}, \dots, m_{tn})$, and let $\tilde{\boldsymbol{p}}_t$ and $\tilde{\boldsymbol{m}}_t$ denote their limiting values. We also let $\boldsymbol{p}_t^*$ and $\boldsymbol{m}_t^*$ denote their unknown true values. Further let $P_0$ represent the true data generating distribution for $\boldsymbol{D}_i$, and let $\mathbb{P}_n$ denote the posterior distribution from a sample of size $n$. We first introduce our assumption on posterior contract rates:

\begin{assumption}[Posterior contraction rates]
\label{asmpt: post contraction rate}
There exist two sequences of numbers $\epsilon_{nt} \rightarrow 0$ and $\epsilon_{ny} \rightarrow 0$, and constants $M_t > 0$ and $M_y > 0$ such that
\begin{enumerate}
	\item[(i)] $\sup\limits_{P_0} E_{P_0} \mathbb{P}_n \bigg( \frac{1}{\sqrt{n}}||\boldsymbol{p}_{t} - \tilde{\boldsymbol{p}}_t||_2 > M_t \epsilon_{nt} \mid \boldsymbol{D} \bigg) \rightarrow 0$,
    \item[(ii)] $\sup\limits_{P_0} E_{P_0} \mathbb{P}_n \bigg( \frac{1}{\sqrt{n}}||\boldsymbol{m}_{t} - \tilde{\boldsymbol{m}}_t||_2 > M_y \epsilon_{ny} \mid \boldsymbol{D} \bigg) \rightarrow 0$.
\end{enumerate}
\end{assumption}
Note that we use $||\boldsymbol{a}||_2 = \sqrt{a_1^2 + \dots + a_n^2}$. This assumption states that the posterior distributions of the propensity score and outcome regression models contract at rates $\epsilon_{nt}$ and $\epsilon_{ny}$, respectively. The standard parametric rate of posterior contraction is $n^{-1/2}$, while slower rates are typically seen for nonparametric Bayesian models or high-dimensional scenarios where the rates depend on the complexity of the model and true data generating process.
\begin{theorem}
Assume positivity, unconfoundedness, SUTVA, Assumption \ref{asmpt: post contraction rate}, and additional regulatory conditions found in Appendix \ref{sec:Assumptions}. Additionally assume that data splitting is used such that half of the data is used for estimating the nuisance parameters, while the other half is used to estimate the CATE. If either $\tilde{\boldsymbol{p}}_t=\boldsymbol{p}_t^*$ or $\tilde{\boldsymbol{m}}_t=\boldsymbol{m}_t^*$, then

$$\sup\limits_{P_0} E_{P_0} \mathbb{P}_n\left(|\Delta(\boldsymbol{D}, \boldsymbol{\Psi})-\tau(\bv)|>M\epsilon_n|\bD\right)\rightarrow0.$$
If both $\tilde{\boldsymbol{p}}_t=\boldsymbol{p}_t^*$ and $\tilde{\boldsymbol{m}}_t=\boldsymbol{m}_t^*$, then $\epsilon_n = \max(n^{-1/2}, \epsilon_{nt} \epsilon_{ny})$. If only one model is correctly specified, then $\epsilon_n$ is equal to the contraction rate of the correctly specified model. 
\label{theorem:posterior contraction}
\end{theorem}

This result importantly implies convergence rates of point estimators such as posterior medians and means (assuming bounded posterior variance). For instance, this result implies that our estimator converges at the $\sqrt{n}$ rate if the product of the contraction rates for the two posterior distributions is $n^{-1/2}$ or smaller. 

\subsection{Theoretical Justification for Variance Estimation} \label{sec:variance estimation}

Now we provide theoretical justification for the variance estimator in (\ref{eqn:VarEstimator}). First, we describe that for general $\Delta(\bPsi, \bD)$ that are functions of both the observed data and unknown parameters, this will provide conservative inference on average. Next we show that for the specific choice of $\Delta(\bPsi, \bD)$ defined in (\ref{eqn:DeltaDef}), this variance estimator is consistent when both the propensity score and outcome regression posterior distributions contract sufficiently fast. 
Throughout this section, let $\widehat{V}$ be the variance estimator defined in (\ref{eqn:VarEstimator}), and let $V$ be the true variance defined by $\text{Var}_{\bD} E_{\bPsi | \bD}[\Delta(\bPsi, \bD)]$. Our goal is to understand whether our variance estimator is conservative in the sense that
\begin{align}
    E_{\bD}\{ \widehat{V} - V \} \gtrapprox 0.
\end{align}
In Appendix \ref{sec:conservativeness}, we prove under sample splitting that this result holds for a particular class of functions $\Delta(\bPsi, \bD)$, and argue why it is expected to hold in general. Additionally, we provide insight into the specific form of the doubly robust estimator and why it is expected that this conservative result will hold for the estimator used in Equation \ref{eqn:DeltaDef}. This result shows that the variance estimator is biased upwards and may lead to conservative inference as our estimates of the variance will be too large. Importantly, this result holds in finite samples or under model misspecification of either the propensity score or outcome regression models. While it is useful to know that our variance estimator tends to be conservative, this could be problematic if the estimated variance is far larger than the true variance, which could lead to meaningless confidence intervals and low power to detect significant treatment effects. In Appendix \ref{sec:upper bound}, we prove that the variance estimator has the following upper bound:

{\footnotesize
\begin{align*}
    &E_{\bD}\{ \widehat{V} \} &\lessapprox 2V-\max\left(\text{Var}_{\bD_{(1)}}\left[E_{\bD_{(2)}} \{ E_{\bPsi | \bD_{(1)}}[\Delta(\bPsi, \bD_{(2)})] \}\right],\text{Var}_{\bD_{(2)}}\left[E_{\bD_{(1)}} \{ E_{\bPsi | \bD_{(1)}}[\Delta(\bPsi, \bD_{(2)})] \}\right]\right)
\end{align*}
}where $\bD_{(1)}$ and $\bD_{(2)}$ are separate splits of the data, which are used to find the posterior distribution of $\boldsymbol{\Psi}$ and estimate the CATE, respectively. This shows that while the variance estimator is conservative, it is generally much smaller than two times the true variance and therefore should not lead to overly wide confidence intervals. 
Additionally, we show that for the doubly robust CATE estimator defined in (\ref{eqn:DeltaDef}), the variance estimator is consistent. 

\begin{theorem}
Let $\Delta(\bPsi, \bD)$ be defined as in (\ref{eqn:DeltaDef}). Assume positivity, unconfoundedness, SUTVA, additional regulatory conditions found in Appendix \ref{sec:Assumptions}, and that data splitting is used. If Assumption \ref{asmpt: post contraction rate} holds with rates $\epsilon_{nt} \leq n^{-{1/4}}$ and $\epsilon_{ny} \leq n^{-{1/4}}$, and both $\tilde{\boldsymbol{p}}_t=\boldsymbol{p}_t^*$ and $\tilde{\boldsymbol{m}}_t=\boldsymbol{m}_t^*$, then $\widehat{V} - V = o_p(n^{-1})$.
\label{theorem:variance}
\end{theorem}

A proof of this can be found in Appendix \ref{sec:consistency}. This states that if both the propensity score and outcome regression models are correctly specified and their posterior distributions contract at rates faster than $n^{-1/4}$, then our variance estimator is consistent for the true variance. These rates can be obtained in high-dimensional settings under sparsity conditions \citep{castillo2015bayesian} or with nonparametric prior distributions under smoothness constraints \citep{van2008rates}. This result, combined with the previous result about the conservative nature of our variance estimator, provides strong justification for using our variance estimation procedure. Additionally, in Appendix \ref{sec:normal approximation} we show our estimator is asymptotically normal under the same conditions as Theorem \ref{theorem:variance}, which justifies the use of a normal approximation for inference. Although this relies on correct specification of both nuisance models, we see empirically in Section \ref{sec:Simulations} that we obtain valid inference even when one model is misspecified.

\section{Simulation Studies}
\label{sec:Simulations}

Here we present the results of simulation studies aimed at evaluating the proposed doubly robust estimator and corresponding variance estimator in nonlinear and high-dimensional modeling scenarios. Throughout this section, we consider the following data-generating process:
$$\bX_i \overset{\text{i.i.d}}{\sim} N(\textbf{0}_p, \textbf{I}_p),\qquad T_i|\bX_i\sim\text{Bernoulli}(p_i),\qquad Y_i|T_i,\bX_i\sim N(\mu_i,1)$$
for $i=1, 2,\cdots,n$ where $\mu_i =\tau(\bV_i)T_i+E(Y(0)|\bX_i)$ and $\bV_i$ consists of the first 10 elements of $\bX_i$. We assume that the true treatment effect, the parameter of interest, is a linear function of covariates:
$$\tau(\bV_i)= 0.3+0.4V_{1i}-0.2V_{2i}+0.7V_{8i}$$
where $V_{ki}$ denotes the $k$-th element of $\bV_{i}$. We explore estimating nonlinear treatment effect functions in Section \ref{sec:Non-linear tau case}. Further, all results for our approaches in this section do not utilize sample splitting or cross-fitting, but a comparison of our approaches with and without sample splitting can be found in Appendix \ref{sec:with and without sample splitting}. 

\subsection{Nonlinear Nuisance Functions}\label{sec:linear}
We first set $p=10$ so that $\bV_i=\bX_i$ for each $i$ and generate data under the following two scenarios for the nuisance functions:
\begin{align*}
\textbf{Linear:}\quad&p_i=\Phi\left(0.3X_{1i}-0.3X_{2i}+0.3X_{3i}-0.3X_{4i}\right)\\
    &E(Y(0)|\bX_i)=0.9X_{1i}-0.6X_{3i}+0.6X_{4i}+0.7X_{6i}\\
\textbf{Nonlinear:}\quad&p_i=\Phi\left(1(X_{1i}>0)-\cos(X_{2i})+0.3|X_{3i}|-\sin(X_{4i})\right)\\
    &E(Y(0)|\bX_i) =\cos(X_{1i})+1(X_{2i}>1)-0.05X_{3i}^3+0.1e^{X_{4i}}+\frac{1}{X_{6i}^2+1}
\end{align*}

We fit the outcome model using a) a Bayesian GLM as in \cite{Gelman2008} (\texttt{DR-Linear}); and b) Bayesian additive regression trees (\texttt{DR-BART}). For both methods, the propensity score is fit with a Bayesian GLM, and inference is performed as in Section \ref{sec:Estimation and inferential strategy}.  In addition to our methods, we use causal forest (\texttt{CF}), BART applied simply to the outcome model (\texttt{BART}), debiased machine learning using both linear regression (\texttt{DML-Linear}) and random forest (\texttt{DML-RF}) for nuisance parameter estimation \citep{Semenova2021}, and Bayesian causal forest of \cite{Hahn2020} (\texttt{BCF}) as our competitors. However, since there is no method in the \texttt{BCF} package for estimating treatment effects at specific locations, \texttt{BCF} is only evaluated in Appendix \ref{sec: training} where we examine $\tau(\bV_i)$ at the observed data locations $\bV_1, \dots, \bV_n$.

\begin{figure}[t]
    \centering
    \includegraphics{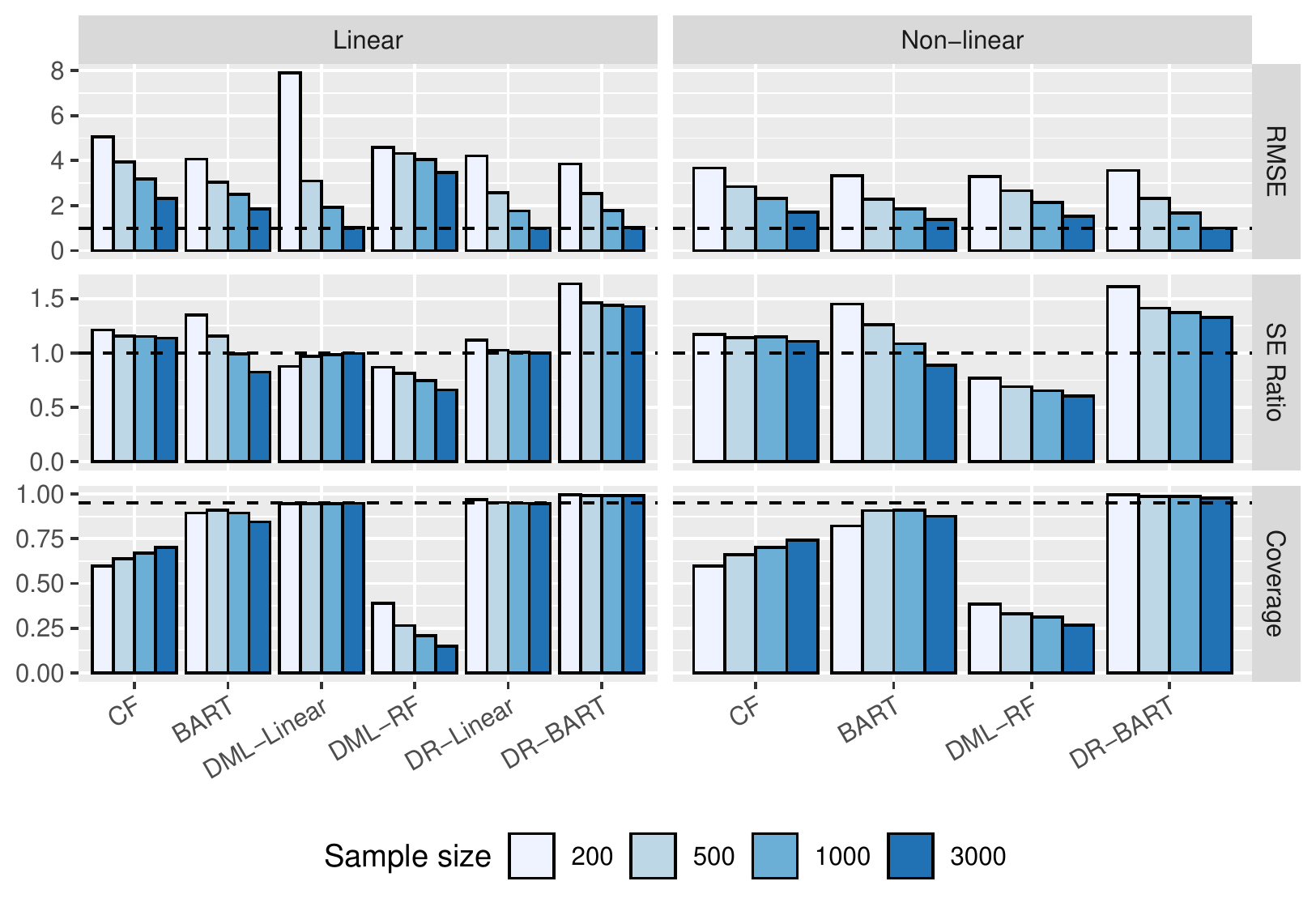}
    \caption{Results for the linear scenario (the first column) and the nonlinear scenario (the second column) for estimating the CATE at randomly chosen locations. The first row shows the scaled root mean squared error, where values are divided by the smallest value within each scenario so that the best RMSE is 1. The second row shows the ratio of average estimated standard errors and Monte Carlo standard errors, and the last row shows the empirical coverage where the dotted line represents the nominal 95\% coverage. }
    \label{fig:linear-nonlinear}
\end{figure}

Our goal is to estimate $\tau(\boldsymbol{V})$ at 100 randomly drawn points from the distribution of $\bV$. We compute empirical 95\% interval coverage, root mean squared error (RMSE) and the ratio of average estimated standard errors and Monte Carlo standard errors. Figure \ref{fig:linear-nonlinear} shows the results of the simulations for both scenarios where the values are averaged over the 100 locations for $\bV$ and 500 simulations. We do not assess the linear methods (\texttt{DML-Linear} and \texttt{DR-Linear}) for the nonlinear scenario since those methods are not expected to perform well in this scenario. 

The proposed methods (\texttt{DR-Linear} and \texttt{DR-BART}) generally have the lowest RMSE among all approaches considered. Regarding inference, the second row of Figure \ref{fig:linear-nonlinear} suggests that our variance estimator is conservative as mentioned in Section \ref{sec:variance estimation}. The \texttt{DR-Linear} approach has a standard error ratio approaching 1 as the sample size grows, which highlights the consistency of our variance estimator. \texttt{DR-BART} appears conservative regardless of sample size, though this is expected, because the contraction rate of BART is far slower than the $n^{-1/4}$ rates required for consistency \citep{rovckova2020posterior}. Note that this stands in stark contrast to the debiased machine learning approaches, which have anti-conservative variance estimates and do not achieve nominal interval coverages. The \texttt{DML-Linear} approach has a consistent estimate of the standard error, though tends to underestimate it in small samples, while the \texttt{DML-RF} approach always underestimates the standard error. The last row shows that our confidence intervals achieve the nominal 95\% coverage for both scenarios. It is notable that \texttt{DR-BART} is the only method that yields valid inference in all settings, and improves on standard \texttt{BART} in terms of RMSE and interval coverage. In short, our method succeeds in constructing valid confidence intervals with small RMSE regardless of the true nuisance functions. We see conservative inference when using BART for nuisance function estimation, but this can easily be reduced by using nonlinear models with faster contraction rates. 

We only considered performance at 100 randomly chosen locations, but we can also look at $\tau(\bV_i)$ at the $n$ observed data locations. We found similar results at the observed data locations, and therefore leave these results to Appendix \ref{sec: training}. One thing to note is that \texttt{BCF} outperforms \texttt{CF} at the observed data locations, but has a slightly larger MSE than the proposed estimators and is anti-conservative for certain sample sizes, though much closer to the nominal level than \texttt{CF}.

\subsection{High-dimensional Nuisance Functions}\label{sec:High-dimensional nuisance functions}
Now we consider the case with the same linear nuisance functions in Section \ref{sec:linear}, but using a high-dimensional $\bX$ and a low-dimensional $\bV$. Specifically, set $n=200, 500, 750$ and $1000$ while $p=2n$ is growing with the sample size. However, we are still interested in inferring $\tau(\bV)$ when $\bV$ consists of the first 10 elements of $\bX$.
\begin{figure}[t]
    \centering
    \includegraphics{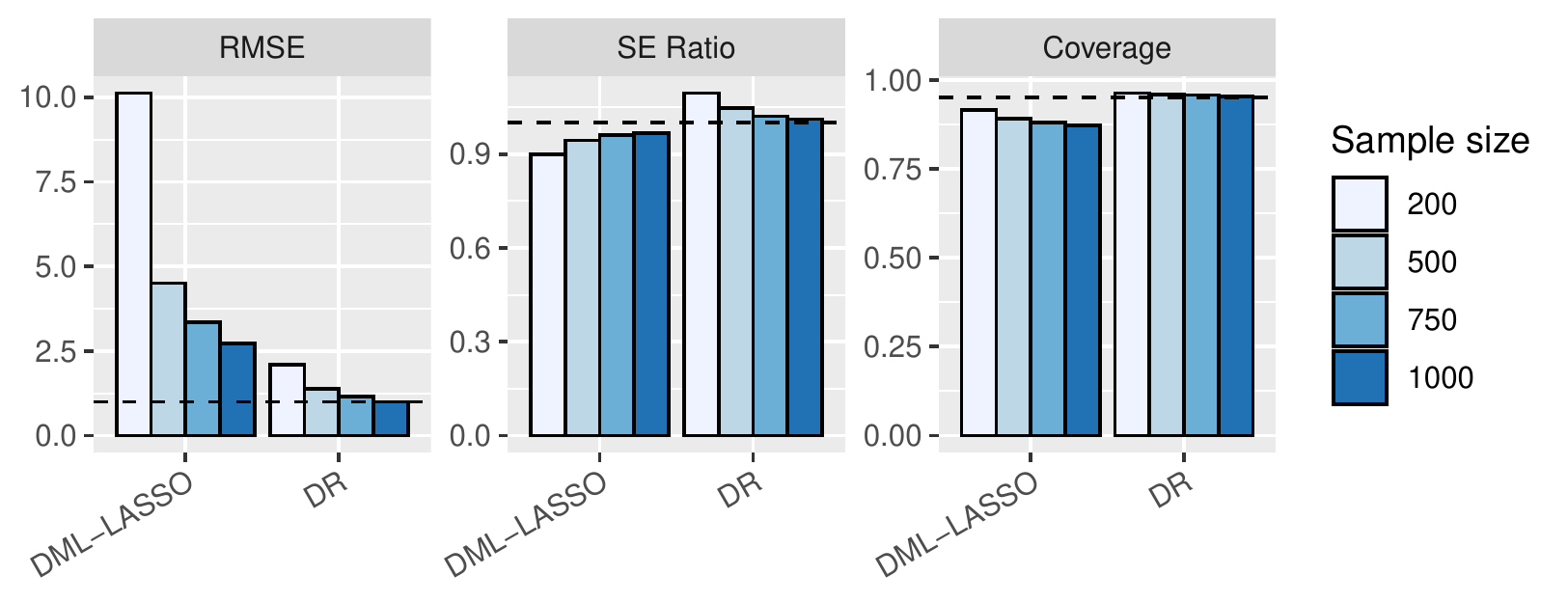}
    \caption{Results for the high-dimensional scenario.}
    \label{fig:HD}
\end{figure}
With high-dimensional $\bX$, we compare our method using spike-and-slab priors to fit the nuisance functions (\texttt{DR}) with debiased machine learning using LASSO (\texttt{DML-LASSO}). The results for the randomly chosen locations are shown in Figure \ref{fig:HD}. Our method (\texttt{DR}) has a smaller RMSE with any sample size and achieves the nominal 95\% coverage in all scenarios. Our variance estimator is slightly conservative for small sample sizes, though converges to the truth as the sample size increases, while the \texttt{DML-LASSO} approach is anti-conservative at each sample size.

\subsection{Nonlinear $\tau(\bV_i)$ Case}\label{sec:Non-linear tau case}
\begin{figure}[t]
    \centering
    \includegraphics{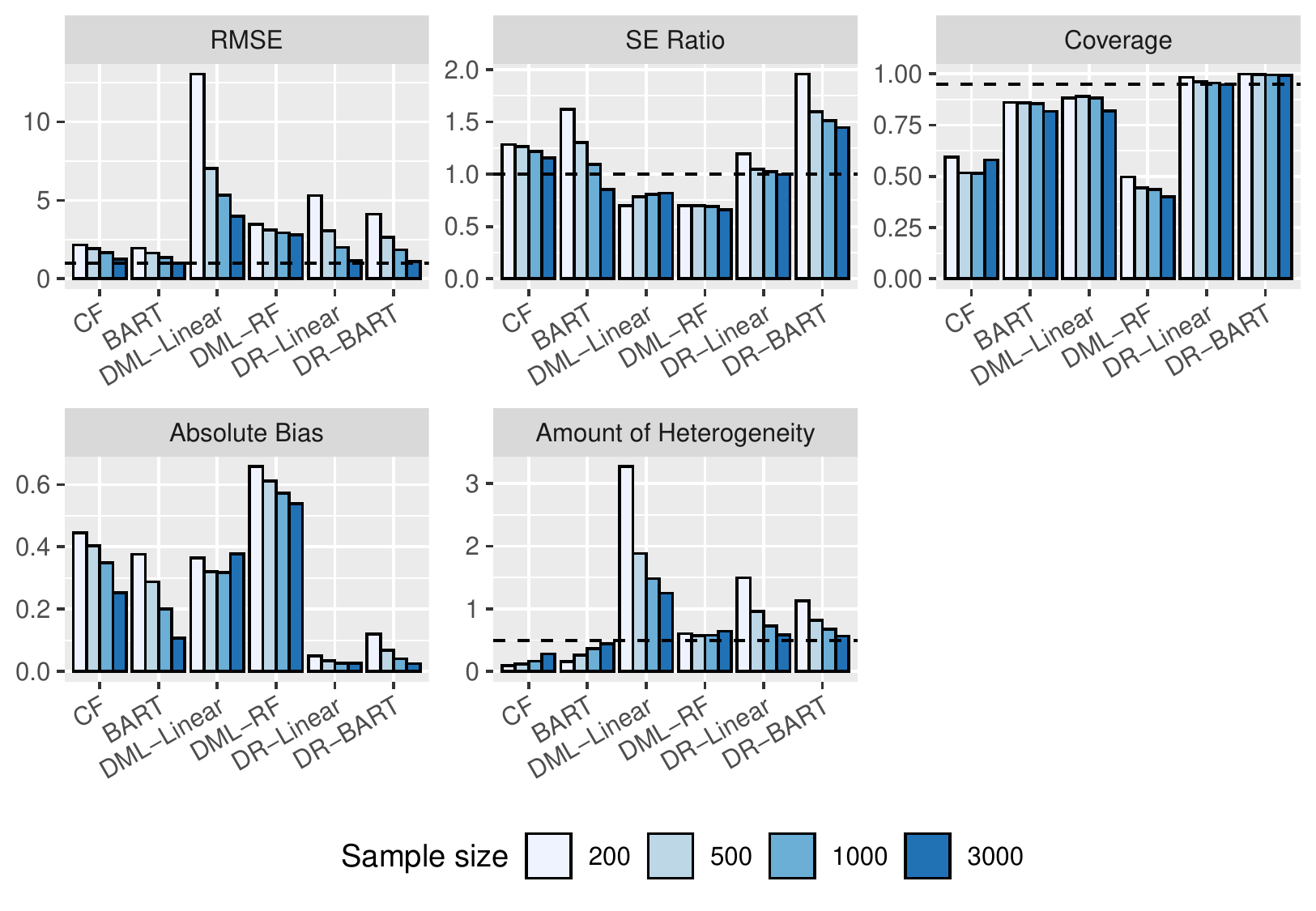}
    \caption{Results for the simulation with non-linear $\tau(\bV)$. The left panel on the second row shows the average absolute bias of each estimator and the right panel shows the average estimated variability of the treatment effects, while the dotted line represents the variability of the true treatment effects.}
    \label{fig:GAM}
\end{figure}

Here we explore empirically the performance of our approach for nonlinear $\tau(\bV)$ functions. In this case, we use a nonlinear parametric method such as natural cubic splines with a fixed degrees of freedom to regress the pseudo outcome on $\bV$. We use the same linear nuisance functions simulation in Section \ref{sec:linear} but replace the linear CATE function with a nonlinear function given by
$$\tau(\bV_i)= 0.3+0.4\cos(V_{1i})-0.2V_{2i}^2+0.7|V_{8i}|,$$
and the results are shown in Figure \ref{fig:GAM}.
Both of our approaches, \texttt{DR-Linear} and \texttt{DR-BART}, are the only ones to achieve interval coverages that are at the nominal 95\% level. \texttt{DR-Linear} achieves exactly nominal coverage while the \texttt{DR-BART} approach is somewhat conservative as expected by our theoretical results, and seen previously in the earlier simulation results. In terms of RMSE, \texttt{CF} and \texttt{BART} have the lowest error when the sample size is small, while our approaches have the lowest error along with \texttt{BART} when the sample size is largest. One might expect that \texttt{CF} and \texttt{BART} should have better coverage rates as they have standard error ratios above 1 and low RMSE, but as we see in the second row of Figure \ref{fig:GAM}, these estimators tend to be biased and overly shrink heterogeneous treatment effect estimates towards an overall homogeneous treatment effect. The right panel of the second row shows the average estimated variability of the treatment effects, which for one particular data set is defined to be $\frac{1}{n-1} \sum_{i=1}^n (\widehat{\tau}(\bV_i) - \frac{1}{n}\sum_{i=1}^n \widehat{\tau}(\bV_i))^2$. We see that \texttt{CF} is substantially underestimating the degree of heterogeneity of the treatment effect, even at large sample sizes, while our proposed estimators are able to capture this heterogeneity. Lastly, we also see that double machine learning estimators are either substantially biased (\texttt{DML-RF}) or have too much estimated heterogeneity (\texttt{DML-Linear}), which leads to worsened performance.

\section{Analysis of Environmental Chemicals}

It is believed that a large portion of disease risk is attributable to differences in one's environment and the types of chemicals or pollutants they are exposed to on a daily basis \citep{patel2014studying}. Therefore, it is critical to understand the effects of environmental exposures on health, and whether these effects vary by subgroups of the population. To help address these questions, we will utilize data from The National Health and Nutrition Examination Survey (NHANES), which is a publicly available data source provided by the United States Centers for Disease Control and Prevention (CDC). We will utilize data that was compiled and provided in \cite{patel2016database}, which contains information from the 1999-2000, 2001-2002, 2003-2004, and 2005-2006 surveys. These data consist of demographic information, physical exam results, laboratory results, and answers from a questionnaire. We will focus on two distinct data sets provided by this database that examine the impacts of two separate environmental exposures: Vitamin B intake, and dialkyl metabolite levels. Vitamin B is a nutrient that contributes to the overall well-being of an individual. Dialkyl metabolite levels can be used to estimate an individual's exposure to organophosphate pesticides, which are known to be toxic to humans. This leads to two distinct analyses on two separate populations from the NHANES data, and we provide details of each, along with results of our analyses below.

\subsection{The Effect of Vitamin B Intake on Triglycerides}

We estimate the treatment effect of vitamin B intake on triglyceride levels, the main constituents of body fat, for 1370 observations using the methods in Section \ref{sec:linear}. Since we do not address continuous treatments in this paper, we first transform the three continuous treatment variables (vitamin B12, serum folate and RBC folate levels) to a single binary treatment in the following manner: each observation is considered treated if at least two out of three treatment levels are greater than the average level of each treatment.

\begin{figure}
    \centering
    \includegraphics[width=6.5 in]{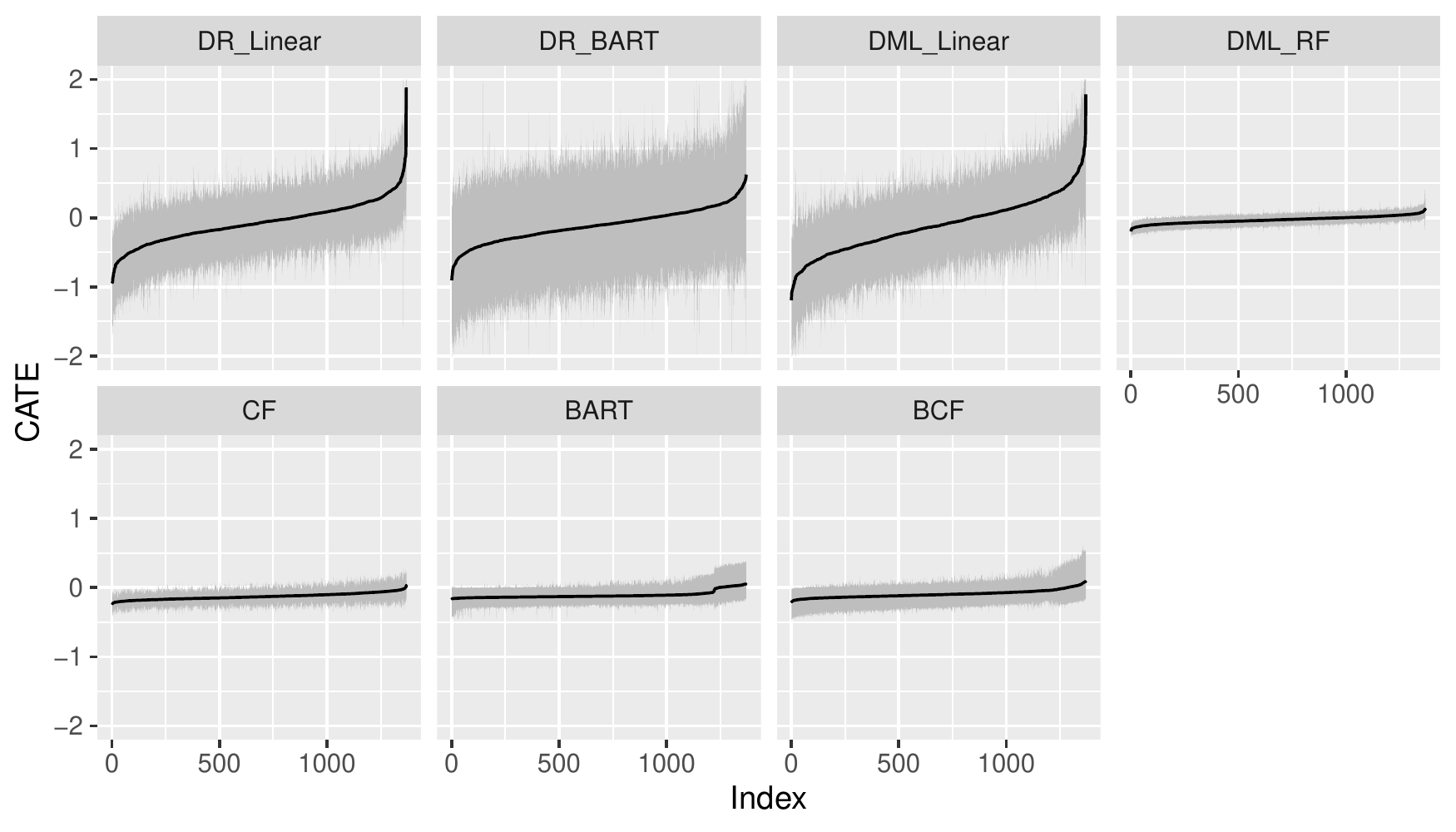}
    \caption{Comparing estimated conditional average treatment effects of vitamin B intake on triglycerides. The solid lines and gray areas represent the ordered CATE estimates and the 95\% confidence intervals, respectively.}
    \label{fig:analysis-low-allX}
\end{figure}

First, we use all variables in the second stage by setting $\bV = \bX$. Figure \ref{fig:analysis-low-allX} shows the estimated $\tau(\bV_i)$ for $i=1, \dots, n$, where the indices of the observations are sorted by estimates of each method. Note these are different from individual treatment effect estimates, and are simply estimates of the CATE evaluated at the observed covariate values $\bV_i$. We see a big difference between the tree-based methods given by \texttt{CF} and \texttt{BCF}, and the remaining approaches. The tree-based approaches show very little variability in these treatment effect estimates, which suggests a homogeneous treatment effect that does not vary by $\bV$. As discussed above in our simulations, we frequently saw the tree-based approaches overly shrinking treatment effect estimates towards homogeneity, and that appears to be occurring here as well. The remaining approaches, including our approaches, lead to far more variability in the estimated conditional average treatment effect indicating some level of variation in the treatment effect. The average treatment effect, which can be estimated by taking the sample average of the conditional average treatment effects, is negative for all approaches considered, which highlights an overall beneficial effect of vitamin B intake that reduces triglyceride levels. 

\subsection{The Effect of Dialkyl Metabolite Levels on HDL Cholesterol}

We now utilize our approach with high-dimensional models to estimate the treatment effect of dialkyl metabolite levels on HDL cholesterol, also known as ``good" cholesterol, for 225 observations with 73 variables. Similar to the vitamin B intake analysis, we convert 6 continuous treatments to a single binary treatment. Due to the large number of variables in $\bX$, we need to pre-specify a subset of variables $\bV$ to investigate heterogeneity by. We select 20 variables that come from demographic information, a physical exam, or a questionnaire as variables to investigate heterogeneity by, since these are more interpretable than the remaining variables which come from laboratory tests.

\begin{figure}
    \centering
    \includegraphics{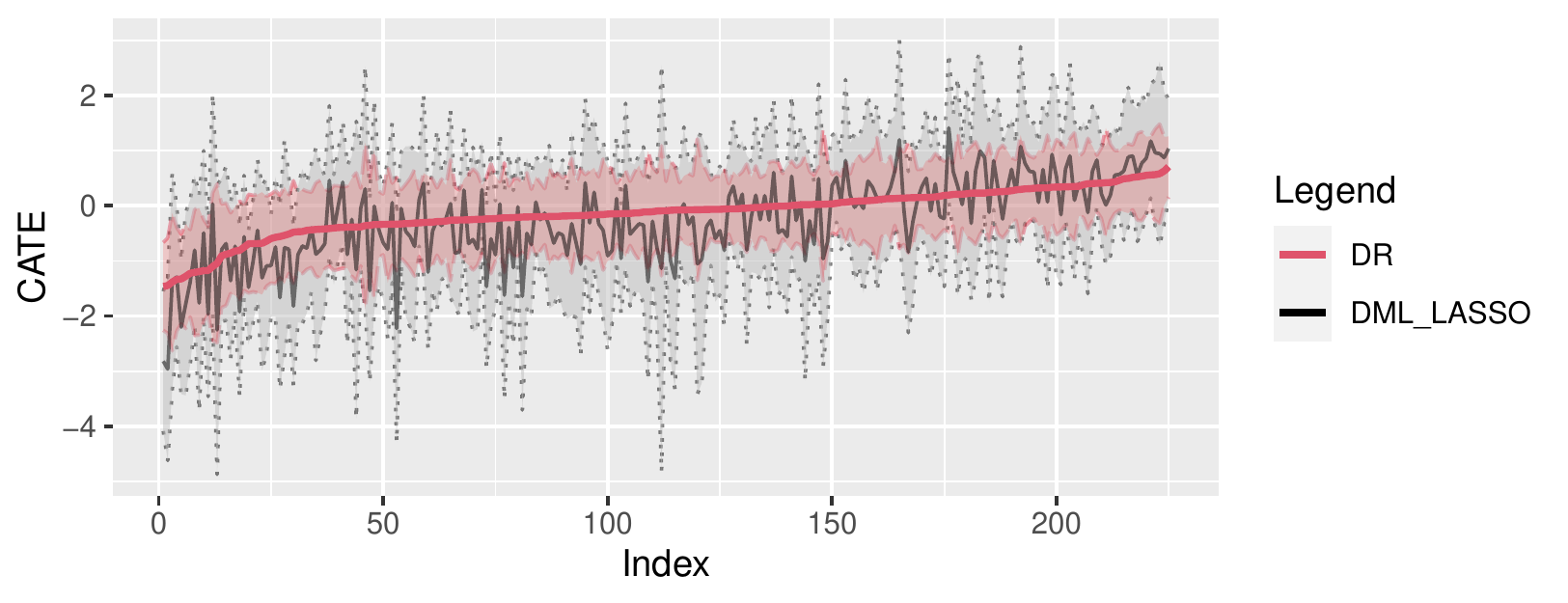}
    \caption{Comparing estimated conditional average treatment effect estimates of dialkyl metabolite levels on HDL cholesterol using 20 covariates. The solid lines and corresponding areas represent the ordered CATE estimates by $\texttt{DR}$, and the 95\% confidence intervals, respectively.}
    \label{fig:analysis-high-allX-est}
\end{figure}

We apply our method (\texttt{DR}) and debiased machine learning with LASSO models (\texttt{DML-LASSO}) to estimate the CATE, and Figure \ref{fig:analysis-high-allX-est} shows the results. Both methods lead to a negative estimated average treatment effect indicating that higher levels of dialkyl metabolites decreases the level of HDL cholesterol. This is expected as dialkyl metabolite levels reflect exposure to toxic pesticides, and lower levels of HDL cholesterol can increase the risk for heart disease or stroke. In terms of heterogeneity, the proposed approach indicates less variability than the \texttt{DML-LASSO} approach as the estimated CATE values are roughly between -1.5 and .5 for our approach, while they are between -3 and 1 for \texttt{DML-LASSO}. Although \texttt{DML-LASSO} gives a larger degree of heterogeneity, this may not stem from true heterogeneity in the treatment effect, but rather from estimation uncertainty. In our high-dimensional simulations of Section \ref{sec:High-dimensional nuisance functions}, we found that \texttt{DML-LASSO} yields a large RMSE compared with the proposed approach.

\begin{figure}
    \centering
    \makebox[\textwidth][c]{
    \includegraphics{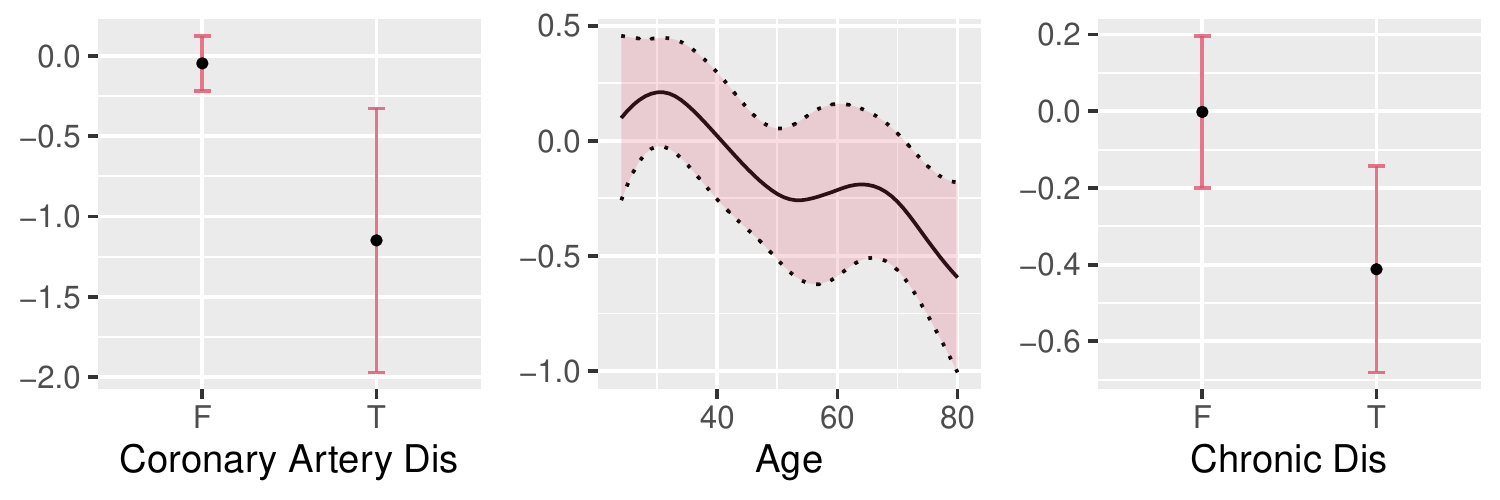}
    }
    \caption{Estimated univariate CATE of dialkyl metabolite levels on the level of HDL cholesterol.}
    
    \label{fig:analysis-high-onevar-all}
\end{figure}

We need not investigate heterogeneity by all 20 variables in $\bV$ jointly, but can rather investigate heterogeneity by one covariate at a time to see if any covariates are strongly associated with the treatment effect. This means we are now estimating $\tau_j(v) = E(Y(1) - Y(0) | V_j = v)$ for $j=1, \dots, p$. This estimand is interesting in its own right as it can provide intuition for any underlying mechanisms that are driving the treatment effect. Given that these are univariate functions, we use cubic splines with a fixed degrees of freedom for estimating the CATE. Figure \ref{fig:analysis-high-onevar-all} shows estimates of these univariate functions for the covariates that have the strongest association with the treatment effect. It appears that dialkyl metabolites have a particularly detrimental effect on the level of HDL cholesterol among subjects with coronary artery disease (CAD), chronic disease, or among older individuals. This leads to an overall understanding that the detrimental impact of dialkyl metabolites is more pronounced among those who are more vulnerable, and may not affect young, healthy individuals to the same degree.

\section{Discussion}

In this paper, we introduced a novel inferential procedure for doubly robust estimators of conditional average treatment effects. We have shown that our variance estimator is consistent when both the propensity score and outcome regression models are correctly specified and contract at sufficiently fast rates, but is conservative in the more difficult settings of finite samples or model misspecification. We have seen empirically that this leads to improved performance in terms of interval coverage compared to existing state-of-the-art approaches, and provides valid inference in difficult high-dimensional or nonparametric modeling situations. The key difference in our approach is that we are able to account for parameter uncertainty by utilizing the posterior distribution of all unknown parameters. This source of uncertainty is difficult to account for in high-dimensional or nonparameteric settings, and is commonly ignored in asymptotic approximations for doubly robust estimators, which can lead to poor performance in finite samples. One might think that this can be solved by bootstrapping competing estimators, though in Appendix \ref{sec: Bootstrap}, we illustrate that the bootstrap can lead to erratic behavior when trying to quantify uncertainty of existing estimators when either high-dimensional or nonparametric models are used. 

There are a number of interesting extensions to be explored in the future. The first is to extend our methods to a continuous treatment, which will allow our method to be applied in a wider range of settings without having to dichotomize treatments as we did in the NHANES data analysis. One could potentially extend either the double machine learning framework \citep{knaus2020double} or the two stage procedure seen in \cite{kennedy2017non} for continuous treatments to allow for heterogeneity of the treatment effect while incorporating our proposed inferential procedure. It would also be interesting to develop theory combining our approach with nonparametric estimates of $\tau(\cdot)$ to reduce assumptions on the CATE. Lastly, we have implicitly assumed throughout that a researcher has a priori knowledge about which elements of $\bX$ are of most interest and should be included in $\bV$. While this is commonly assumed in models estimating treatment effect heterogeneity with a high-dimensional $\bX$, it would be interesting to develop a data-driven approach to selecting the variables to include in $\bV$ that still permits valid inference on the CATE.

\bibliographystyle{apalike}
\bibliography{ref2.bib}

\begin{thebibliography}{}

\bibitem[Abrevaya et~al., 2015]{Abrevaya2015}
Abrevaya, J., Hsu, Y.-C., and Lieli, R.~P. (2015).
\newblock {Estimating Conditional Average Treatment Effects}.
\newblock {\em Journal of Business {\&} Economic Statistics}, 33(4):485--505.

\bibitem[Antonelli and Cefalu, 2020]{antonelli2020averaging}
Antonelli, J. and Cefalu, M. (2020).
\newblock Averaging causal estimators in high dimensions.
\newblock {\em Journal of Causal Inference}, 8(1):92--107.

\bibitem[Antonelli et~al., 2018]{antonelli2018doubly}
Antonelli, J., Cefalu, M., Palmer, N., and Agniel, D. (2018).
\newblock Doubly robust matching estimators for high dimensional confounding
  adjustment.
\newblock {\em Biometrics}, 74(4):1171--1179.

\bibitem[Antonelli et~al., 2022]{antonelli2022causal}
Antonelli, J., Papadogeorgou, G., and Dominici, F. (2022).
\newblock Causal inference in high dimensions: A marriage between bayesian
  modeling and good frequentist properties.
\newblock {\em Biometrics}, 78(1):100--114.

\bibitem[Antonelli et~al., 2019]{antonelli2019high}
Antonelli, J., Parmigiani, G., and Dominici, F. (2019).
\newblock High-dimensional confounding adjustment using continuous spike and
  slab priors.
\newblock {\em Bayesian analysis}, 14(3):805.

\bibitem[Athey and Imbens, 2016]{Athey2016}
Athey, S. and Imbens, G. (2016).
\newblock {Recursive partitioning for heterogeneous causal effects}.
\newblock {\em Proceedings of the National Academy of Sciences of the United
  States of America}, 113(27):7353--7360.

\bibitem[Avagyan and Vansteelandt, 2021]{avagyan2021high}
Avagyan, V. and Vansteelandt, S. (2021).
\newblock High-dimensional inference for the average treatment effect under
  model misspecification using penalized bias-reduced double-robust estimation.
\newblock {\em Biostatistics \& Epidemiology}, pages 1--18.

\bibitem[Bang and Robins, 2005]{bang2005doubly}
Bang, H. and Robins, J.~M. (2005).
\newblock Doubly robust estimation in missing data and causal inference models.
\newblock {\em Biometrics}, 61(4):962--973.

\bibitem[Belloni et~al., 2014]{belloni2014high}
Belloni, A., Chernozhukov, V., and Hansen, C. (2014).
\newblock High-dimensional methods and inference on structural and treatment
  effects.
\newblock {\em Journal of Economic Perspectives}, 28(2):29--50.

\bibitem[Benkeser et~al., 2017]{benkeser2017doubly}
Benkeser, D., Carone, M., Laan, M. V.~D., and Gilbert, P. (2017).
\newblock Doubly robust nonparametric inference on the average treatment
  effect.
\newblock {\em Biometrika}, 104(4):863--880.

\bibitem[Castillo et~al., 2015]{castillo2015bayesian}
Castillo, I., Schmidt-Hieber, J., and Van~der Vaart, A. (2015).
\newblock Bayesian linear regression with sparse priors.
\newblock {\em The Annals of Statistics}, 43(5):1986--2018.

\bibitem[Chernozhukov et~al., 2018]{Chernozhukov2018}
Chernozhukov, V., Chetverikov, D., Demirer, M., Duflo, E., Hansen, C., Newey,
  W., and Robins, J. (2018).
\newblock {Double/debiased machine learning for treatment and structural
  parameters}.
\newblock {\em Econometrics Journal}, 21(1):C1--C68.

\bibitem[Chipman et~al., 2012]{Chipman2012}
Chipman, H.~A., George, E.~I., and McCulloch, R.~E. (2012).
\newblock {BART: Bayesian additive regression trees}.
\newblock {\em Annals of Applied Statistics}, 4(1):266--298.

\bibitem[Dorie et~al., 2019]{Dorie2019}
Dorie, V., Hill, J., Shalit, U., Scott, M., and Cervone, D. (2019).
\newblock {Automated versus Do-It-Yourself Methods for Causal Inference:
  Lessons Learned from a Data Analysis Competition 1}.
\newblock {\em Statistical Science}, 34(1):43--68.

\bibitem[Dukes et~al., 2020]{dukes2020doubly}
Dukes, O., Avagyan, V., and Vansteelandt, S. (2020).
\newblock Doubly robust tests of exposure effects under high-dimensional
  confounding.
\newblock {\em Biometrics}, 76(4):1190--1200.

\bibitem[Dukes et~al., 2021]{dukes2021doubly}
Dukes, O., Vansteelandt, S., and Whitney, D. (2021).
\newblock On doubly robust inference for double machine learning.
\newblock {\em arXiv preprint arXiv:2107.06124}.

\bibitem[Efron and Tibshirani, 1994]{efron1994introduction}
Efron, B. and Tibshirani, R.~J. (1994).
\newblock {\em An introduction to the bootstrap}.
\newblock CRC press.

\bibitem[El~Karoui and Purdom, 2018]{el2018can}
El~Karoui, N. and Purdom, E. (2018).
\newblock Can we trust the bootstrap in high-dimensions? the case of linear
  models.
\newblock {\em The Journal of Machine Learning Research}, 19(1):170--235.

\bibitem[Fan et~al., 2020]{Fan2020}
Fan, Q., Hsu, Y.~C., Lieli, R.~P., and Zhang, Y. (2020).
\newblock {Estimation of Conditional Average Treatment Effects With
  High-Dimensional Data}.
\newblock {\em Journal of Business and Economic Statistics}.

\bibitem[Farrell, 2015]{Farrell2015}
Farrell, M.~H. (2015).
\newblock {Robust inference on average treatment effects with possibly more
  covariates than observations}.
\newblock {\em Journal of Econometrics}, 189(1):1--23.

\bibitem[Gelman et~al., 2008]{Gelman2008}
Gelman, A., Jakulin, A., Pittau, M.~G., and Su, Y.~S. (2008).
\newblock {A weakly informative default prior distribution for logistic and
  other regression models}.
\newblock {\em Annals of Applied Statistics}, 2(4):1360--1383.

\bibitem[Hahn et~al., 2020]{Hahn2020}
Hahn, P.~R., Murray, J.~S., and Carvalho, C.~M. (2020).
\newblock {Bayesian regression tree models for causal inference:
  Regularization, confounding, and heterogeneous effects (with discussion)}.
\newblock {\em Bayesian Analysis}, 15(3):965--1056.

\bibitem[Hill, 2011]{Hill2011}
Hill, J.~L. (2011).
\newblock {Bayesian Nonparametric Modeling for Causal Inference}.
\newblock {\em Journal of Computational and Graphical Statistics},
  20(1):217--240.

\bibitem[Kennedy, 2020]{kennedy2020optimal}
Kennedy, E.~H. (2020).
\newblock Optimal doubly robust estimation of heterogeneous causal effects.
\newblock {\em arXiv preprint arXiv:2004.14497}.

\bibitem[Kennedy et~al., 2017]{kennedy2017non}
Kennedy, E.~H., Ma, Z., McHugh, M.~D., and Small, D.~S. (2017).
\newblock Non-parametric methods for doubly robust estimation of continuous
  treatment effects.
\newblock {\em Journal of the Royal Statistical Society: Series B (Statistical
  Methodology)}, 79(4):1229--1245.

\bibitem[Knaus, 2020]{knaus2020double}
Knaus, M.~C. (2020).
\newblock Double machine learning based program evaluation under
  unconfoundedness.
\newblock {\em arXiv preprint arXiv:2003.03191}.

\bibitem[K{\"{u}}nzel et~al., 2019]{Kunzel2019}
K{\"{u}}nzel, S.~R., Sekhon, J.~S., Bickel, P.~J., and Yu, B. (2019).
\newblock {Metalearners for estimating heterogeneous treatment effects using
  machine learning}.
\newblock {\em Proceedings of the National Academy of Sciences of the United
  States of America}, 116(10):4156--4165.

\bibitem[Lee et~al., 2020]{lee2020causal}
Lee, K., Bargagli-Stoffi, F.~J., and Dominici, F. (2020).
\newblock Causal rule ensemble: Interpretable inference of heterogeneous
  treatment effects.
\newblock {\em arXiv preprint arXiv:2009.09036}.

\bibitem[Lee et~al., 2021]{Lee2021}
Lee, K., Small, D.~S., and Dominici, F. (2021).
\newblock Discovering heterogeneous exposure effects using randomization
  inference in air pollution studies.
\newblock {\em Journal of the American Statistical Association}, 116:569--580.

\bibitem[Lee et~al., 2017]{Lee2017}
Lee, S., Okui, R., and Whang, Y.-J. (2017).
\newblock {Doubly robust uniform confidence band for the conditional average
  treatment effect function}.
\newblock {\em Journal of Applied Econometrics}, 32(7):1207--1225.

\bibitem[Little and Rubin, 2000]{little2000causal}
Little, R.~J. and Rubin, D.~B. (2000).
\newblock Causal effects in clinical and epidemiological studies via potential
  outcomes: concepts and analytical approaches.
\newblock {\em Annual review of public health}, 21(1):121--145.

\bibitem[Nie and Wager, 2021]{Nie2021}
Nie, X. and Wager, S. (2021).
\newblock {Quasi-oracle estimation of heterogeneous treatment effects}.
\newblock {\em Biometrika}, 108(2):299--319.

\bibitem[Ning et~al., 2020]{ning2020robust}
Ning, Y., Sida, P., and Imai, K. (2020).
\newblock Robust estimation of causal effects via a high-dimensional covariate
  balancing propensity score.
\newblock {\em Biometrika}, 107(3):533--554.

\bibitem[Patel and Ioannidis, 2014]{patel2014studying}
Patel, C.~J. and Ioannidis, J.~P. (2014).
\newblock Studying the elusive environment in large scale.
\newblock {\em Jama}, 311(21):2173--2174.

\bibitem[Patel et~al., 2016]{patel2016database}
Patel, C.~J., Pho, N., McDuffie, M., Easton-Marks, J., Kothari, C., Kohane,
  I.~S., and Avillach, P. (2016).
\newblock A database of human exposomes and phenomes from the us national
  health and nutrition examination survey.
\newblock {\em Scientific data}, 3(1):1--10.

\bibitem[Powers et~al., 2018]{Powers2018}
Powers, S., Qian, J., Jung, K., Schuler, A., Shah, N.~H., Hastie, T., and
  Tibshirani, R. (2018).
\newblock {Some methods for heterogeneous treatment effect estimation in high
  dimensions}.
\newblock {\em Statistics in Medicine}, 37(11):1767--1787.

\bibitem[Ro{\v{c}}kov{\'a} and van~der Pas, 2020]{rovckova2020posterior}
Ro{\v{c}}kov{\'a}, V. and van~der Pas, S. (2020).
\newblock Posterior concentration for bayesian regression trees and forests.
\newblock {\em The Annals of Statistics}, 48(4):2108--2131.

\bibitem[Scharfstein et~al., 1999]{scharfstein1999adjusting}
Scharfstein, D.~O., Rotnitzky, A., and Robins, J.~M. (1999).
\newblock Adjusting for nonignorable drop-out using semiparametric nonresponse
  models.
\newblock {\em Journal of the American Statistical Association},
  94(448):1096--1120.

\bibitem[Semenova and Chernozhukov, 2021]{Semenova2021}
Semenova, V. and Chernozhukov, V. (2021).
\newblock {Debiased machine learning of conditional average treatment effects
  and other causal functions}.
\newblock {\em The Econometrics Journal}, 24(2):264--289.

\bibitem[Tan, 2020a]{tan2020model}
Tan, Z. (2020a).
\newblock Model-assisted inference for treatment effects using regularized
  calibrated estimation with high-dimensional data.
\newblock {\em The Annals of Statistics}, 48(2):811--837.

\bibitem[Tan, 2020b]{tan2020regularized}
Tan, Z. (2020b).
\newblock Regularized calibrated estimation of propensity scores with model
  misspecification and high-dimensional data.
\newblock {\em Biometrika}, 107(1):137--158.

\bibitem[Van~der Laan, 2014]{van2014targeted}
Van~der Laan, M.~J. (2014).
\newblock Targeted estimation of nuisance parameters to obtain valid
  statistical inference.
\newblock {\em The international journal of biostatistics}, 10(1):29--57.

\bibitem[van~der Vaart, 1998]{Vaart1998}
van~der Vaart, A.~W. (1998).
\newblock {\em Asymptotic Statistics}.
\newblock Cambridge University Press.

\bibitem[van~der Vaart and van Zanten, 2008]{van2008rates}
van~der Vaart, A.~W. and van Zanten, J.~H. (2008).
\newblock Rates of contraction of posterior distributions based on gaussian
  process priors.
\newblock {\em The Annals of Statistics}, 36(3):1435--1463.

\bibitem[Wager and Athey, 2018]{Wager2018}
Wager, S. and Athey, S. (2018).
\newblock {Estimation and Inference of Heterogeneous Treatment Effects using
  Random Forests}.
\newblock {\em Journal of the American Statistical Association},
  113(523):1228--1242.

\bibitem[Wang and Rudin, 2022]{Wang2022}
Wang, T. and Rudin, C. (2022).
\newblock Causal rule sets for identifying subgroups with enhanced treatment
  effects.
\newblock {\em INFORMS Journal on Computing}, 34:1626--1643.

\end{thebibliography}

\appendix

\section{Proof of consistency of variance estimator}\label{sec:consistency}

We will focus on the estimation of $E(Y(t) | \bV = \bv)$ where $\bV$ is an $n \times q$ matrix of covariates that we believe may modify the treatment effect. Note here that $\bV$ is a subset of the columns of $\bX$, i.e. $q \leq p$ and we are assuming that $q$ is finite-dimensional and does not grow with $n$. Of most importance is that $\frac{1}{n} \bV^T \bV \rightarrow \Sigma_V$ for some matrix $\Sigma_V$ and that $(\bV^T \bV)^{-1}$ exists and is unique. The extension of our result to the CATE defined by $E(Y(1) - Y(0) | \bV = \bv)$ is trivial once we have the result for both of the potential outcomes separately. Throughout, the $*$ and $\sim$ superscript are used to denote the true and limiting value of that quantity, respectively. As in the paper, we will denote the pseudo-outcome as
$$Z_i = \frac{1(T_i = t)}{p_{ti}} (Y_i - m_{ti}) + m_{ti},$$
and we will regularly make use of the fact that this can be decomposed as follows:
\begin{align*}
    Z_i &= \frac{1(T_i = t)}{p_{ti}} (Y_i - m_{ti}) + m_{ti} \\
    &= (m_{ti} - m_{ti}^*) \Bigg(1 - \frac{1(T_i = t)}{p_{ti}^*} \Bigg) \\
    &+ \frac{1(T_i = t)(p_{ti} - p_{ti}^*) (m_{ti}^* - Y_i)}{p_{ti}p_{ti}^*} \\
    &+ \frac{1(T_i = t)(p_{ti} - p_{ti}^*) (m_{ti} - m_{ti}^*)}{p_{ti}p_{ti}^*} \\
    &+ \frac{1(T_i = t)}{p_{ti}^*} (Y_i - m_{ti}^*) + m_{ti}^* \\
    &= A_{1i} + A_{2i} + A_{3i} + B_i
\end{align*}

Note that each of these terms is a function of both data $\boldsymbol{D}$ and parameters $\boldsymbol{\Psi}$ though we will suppress this dependence for brevity, i.e. we adopt the convention that $A_{1i} = A_{1}(\boldsymbol{D}_i, \boldsymbol{\Psi})$, etc. We will also define $n \times 1$ vectors of these quantities evaluated at each of the $n$ data points, which we will denote by $\boldsymbol{Z}$,$\boldsymbol{A}_1$, $\boldsymbol{A}_2$, $\boldsymbol{A}_3$, and, $\boldsymbol{B}$. Given a parameter vector $\bPsi$, an estimator of $E(Y(t) | \bV = \bv)$ can then be defined as
$$\Delta(\boldsymbol{D}, \boldsymbol{\Psi}) = \bv (\bV^T \bV)^{-1} \bV^T \bZ = \bv (\bV^T \bV)^{-1} \bV^T (\bAone + \bAtwo + \bAthree + \bB).$$
In our manuscript, our estimator is defined to be the posterior mean of this quantity, i.e. $E_{\bPsi | \bD} \Big[ \Delta(\boldsymbol{D}, \boldsymbol{\Psi}) \Big]$, and our variance estimator is given by 
$$\widehat{V} = \text{Var}_{\boldsymbol{D^{(m)}}} \{ E_{\boldsymbol{\Psi} \vert \boldsymbol{D}} [\Delta(\boldsymbol{D^{(m)}}, \boldsymbol{\Psi})] \} + \text{Var}_{\boldsymbol{\Psi} \vert \boldsymbol{D}} [\Delta(\boldsymbol{D}, \boldsymbol{\Psi})].$$
The true variance that we wish to estimate is given by $V = \text{Var}_{\boldsymbol{D}} \{ E_{\boldsymbol{\Psi} \vert \boldsymbol{D}} [\Delta(\boldsymbol{D}, \boldsymbol{\Psi})] \}$. Our goal will be to show that $\widehat{V} - V = o_p(n^{-1})$ when both models are correctly specified and contract at sufficiently fast rates. To do this, we will detail the asymptotic behavior of three components separately. First, we will show that $\text{Var}_{\boldsymbol{\Psi} \vert \boldsymbol{D}} [\Delta(\boldsymbol{D}, \boldsymbol{\Psi})] = o_p(n^{-1})$. This implies that the posterior variability of the estimator is negligible asymptotically. Next, we show that $\text{Var}_{\boldsymbol{D^{(m)}}} \{ E_{\boldsymbol{\Psi} \vert \boldsymbol{D}} [\Delta(\boldsymbol{D^{(m)}}, \boldsymbol{\Psi})] \} = \text{Var}_{\boldsymbol{D}} \Big[ \bv (\bV^T \bV)^{-1} \bV^T \bB \Big] + o_p(n^{-1})$, and that $\text{Var}_{\boldsymbol{D}} \{ E_{\boldsymbol{\Psi} \vert \boldsymbol{D}} [\Delta(\boldsymbol{D}, \boldsymbol{\Psi})] \} = \text{Var}_{\boldsymbol{D}}\Big[ \bv (\bV^T \bV)^{-1} \bV^T \bB \Big] + o_p(n^{-1})$. These three results imply that our variance estimator is consistent as the difference between our estimate and the true variance is $o_p(n^{-1})$ and is asymptotically negligible. Before showing the proof, we now detail assumptions required on the data generating process and the posterior distributions of both the treatment and outcome models. 

\subsection{Preliminary Assumptions}
\label{sec:Assumptions}

We introduce the regularity assumptions in this section. While Assumption \ref{asmpt: moments} is required only for the proof of consistency of the proposed variance estimator in Appendix \ref{sec:consistency}, we will use Assumption \ref{asmpt: DGP}-\ref{asmpt: Appendix contraction rates} for all theoretical proofs in Appendix \ref{sec:consistency}-\ref{sec:normal approximation}.
\begin{assumption}[Data generating process]\label{asmpt: DGP}\leavevmode

\begin{enumerate}
	\item[(a)] $\left\{(Y_i, T_i, \boldsymbol{X}_i) \right\}_{i=1}^n$ are i.i.d samples from $P_0$
    \item[(b)] The covariates $X_j$ have bounded support, in that there exists $K_x < \infty$ such that $|X_j| < K_x$ with probability 1 for all $j$.
    \item[(c)] $\sup\limits_{P_0} E_{P_0}((Y - m_{ti}^*)^2) \leq K_y < \infty$.
\end{enumerate}
\end{assumption}

Assumption (a) and (b) are standard and likely to be satisfied in real applications as most variables are naturally bounded. Assumption (c) ensures that the residual variance of the outcome is bounded, which again should be satisfied in most applications.

\begin{assumption}[Bounds on the posterior distribution]\label{asmpt: posterior bounds}\leavevmode
\begin{enumerate}
	\item[(a)] $\sup\limits_{P_0} E_{P_0} \text{Var}_n \bigg( \frac{p_{ti} - p_{ti}^*}{p_{ti}} \vert \boldsymbol{D}_i \bigg) \leq K_p < \infty$
    \item[(b)] $\sup\limits_{P_0} E_{P_0} \text{Var}_n \bigg( m_{ti} - m_{ti}^* \vert \boldsymbol{D}_i \bigg) \leq K_m < \infty$
\end{enumerate}
\end{assumption}

Assumption (a) effectively states that the posterior distribution of $p_{ti}$ does not assign mass to neighborhoods of 0, while assumption (b) states that the difference between the true conditional mean of the outcome and the corresponding posterior is bounded. These are mild assumptions in general, and should hold under reasonable models and posterior distributions. 

\begin{assumption}[Posterior contraction rates]\label{asmpt: Appendix contraction rates}
There exist two sequence of numbers $\epsilon_{nt} \rightarrow 0$ and $\epsilon_{ny} \rightarrow 0$, and constants $M_t > 0$ and $M_y > 0$ such that
\begin{enumerate}
	\item[(a)] $\sup\limits_{P_0} E_{P_0} \mathbb{P}_n \bigg( \frac{1}{\sqrt{n}}||\boldsymbol{p}_{t} - \tilde{\boldsymbol{p}}_{t}||_2 > M_t \epsilon_{nt} \vert \boldsymbol{D} \bigg) \rightarrow 0$, and
    \item[(b)] $\sup\limits_{P_0} E_{P_0} \mathbb{P}_n \bigg( \frac{1}{\sqrt{n}}||\boldsymbol{m}_{t} - \tilde{\boldsymbol{m}}_{t}||_2 > M_y \epsilon_{ny} \vert \boldsymbol{D} \bigg) \rightarrow 0$,
\end{enumerate}
where $||v||_2 = \sqrt{v_1^2 + \dots + v_n^2}$.
\end{assumption}

Assumption (a) and (b) state that the posterior distribution of the treatment and outcome models contract at rates $\epsilon_{nt}$ and $\epsilon_{ny}$, respectively. For instance, in fully parametric settings, we would expect these models to contract at rates of $n^{-1/2}$, which is analogous to the frequentist setting of $\sqrt{n}-$convergence.  Posterior contraction rates are dependent on the models chosen and usually rely on their own set of assumptions, such as conditions on the design matrix $\boldsymbol{X}$, smoothness, or sparsity. We will not detail these assumptions here, as our goal will be to show what happens to our variance estimator if the treatment and outcome models contract at specific rates. 

Lastly, we need to make assumptions on specific moments of the posterior distribution, in particular that they converge sufficiently fast to zero. 

\begin{assumption}[Moments of the posterior distribution]\label{asmpt: moments}\leavevmode
\begin{enumerate}
	\item $$\sum_{a=1}^n \sum_{b=1}^n \sqrt{\text{E}_{\boldsymbol{\Psi} \vert \boldsymbol{D}} \bigg[(m_{ta} - \tilde{m}_{ta})^2 (m_{tb} - \tilde{m}_{tb})^2 \bigg]} \sqrt{\text{E}_{\boldsymbol{\Psi} \vert \boldsymbol{D}} \bigg[(p_{ta} - \tilde{p}_{ta})^2 (p_{tb} - \tilde{p}_{tb})^2 \bigg]} = o_p(n)$$
	\item \begin{align*}
	    \sum_{a=1}^n \sum_{b=1}^n &\sqrt{\text{E}_{\boldsymbol{\Psi} \vert \boldsymbol{D}} \bigg[(m_{ta} - \tilde{m}_{ta})^2 \bigg] \text{E}_{\boldsymbol{\Psi} \vert \boldsymbol{D}} \bigg[(p_{ta} - \tilde{p}_{ta})^2 \bigg]}\\ &\qquad\times\sqrt{\text{E}_{\boldsymbol{\Psi} \vert \boldsymbol{D}} \bigg[(m_{tb} - \tilde{m}_{tb})^2 \bigg] \text{E}_{\boldsymbol{\Psi} \vert \boldsymbol{D}} \bigg[(p_{tb} - \tilde{p}_{tb})^2 \bigg]} = o_p(n)
	\end{align*}
	\item $$\sum_{b=1}^n   \text{E}_{\bD} \Bigg\{ \text{E}_{\boldsymbol{\Psi} \vert \boldsymbol{D}} \Bigg[ (m_{ta} - \tilde{m}_{ta})  (m_{tb} - \tilde{m}_{tb}) \Bigg]^2 \Bigg\} = o(1)$$
	\item $$\sum_{b=1}^n   \text{E}_{\bD} \Bigg\{ \text{E}_{\boldsymbol{\Psi} \vert \boldsymbol{D}} \Bigg[ (p_{ta} - \tilde{p}_{ta})  (p_{tb} - \tilde{p}_{tb}) \Bigg]^2 \Bigg\} = o(1)$$
	\item $$\sum_{b=1}^n   \text{E}_{\bD} \Bigg\{ \text{E}_{\boldsymbol{\Psi} \vert \boldsymbol{D}} \Bigg[ (m_{ta} - \tilde{m}_{ta}) \Bigg]^2  \text{E}_{\boldsymbol{\Psi} \vert \boldsymbol{D}} \Bigg[ (m_{tb} - \tilde{m}_{tb}) \Bigg]^2 \Bigg\} = o(1)$$
	\item $$\sum_{b=1}^n   \text{E}_{\bD} \Bigg\{ \text{E}_{\boldsymbol{\Psi} \vert \boldsymbol{D}} \Bigg[ (p_{ta} - \tilde{p}_{ta}) \Bigg]^2  \text{E}_{\boldsymbol{\Psi} \vert \boldsymbol{D}} \Bigg[ (p_{tb} - \tilde{p}_{tb}) \Bigg]^2 \Bigg\} = o(1)$$
\end{enumerate}
\end{assumption}

 These assumptions are similar to assumptions typically made in the high-dimensional and semiparametric causal inference literature \cite{Farrell2015} that assume the convergence rates of the propensity score or outcome regression models are slightly faster than $n^{-{1/4}}$. These expectations are all closely related to the posterior contraction of the propensity score and outcome regression models, and are expected to hold when these two models contract at the $n^{-1/4}$ or faster rates that are assumed in Assumption \ref{asmpt: Appendix contraction rates}. 

\subsection{Proof}

\subsubsection{Asymptotic Behavior of $\text{Var}_{\boldsymbol{\Psi} \vert \boldsymbol{D}} [\Delta(\boldsymbol{D}, \boldsymbol{\Psi})]$}

We will begin by detailing the asymptotic behavior of each of the four components of $\Delta(\boldsymbol{D}, \boldsymbol{\Psi})$ separately. We will analyze $\text{Var}_{\boldsymbol{\Psi} \vert \boldsymbol{D}} [\bv (\bV^T \bV)^{-1} \bV^T \bAone]$, $\text{Var}_{\boldsymbol{\Psi} \vert \boldsymbol{D}} [\bv (\bV^T \bV)^{-1} \bV^T \bAtwo]$, $\text{Var}_{\boldsymbol{\Psi} \vert \boldsymbol{D}} [\bv (\bV^T \bV)^{-1} \bV^T \bAthree]$, and $\text{Var}_{\boldsymbol{\Psi} \vert \boldsymbol{D}} [\bv (\bV^T \bV)^{-1} \bV^T \bB]$. If each of these four components separately are $o_p(n^{-1})$, then the entire variance must be of the same order, because all relevant covariances will be of the same or lower order. The first observation to make is that $\bv (\bV^T \bV)^{-1} \bV^T \bB$ contains no unknown parameters and therefore $\text{Var}_{\boldsymbol{\Psi} \vert \boldsymbol{D}} [\bv (\bV^T \bV)^{-1} \bV^T \bB] = 0$. \\
\\
$\bAone$: \\
\begin{align*}
    \text{Var}_{\boldsymbol{\Psi} \vert \boldsymbol{D}} [\bv (\bV^T \bV)^{-1} \bV^T \bAone] &\leq \text{E}_{\boldsymbol{\Psi} \vert \boldsymbol{D}} [\bv (\bV^T \bV)^{-1} \bV^T \bAone \bAone^T \bV (\bV^T \bV)^{-1} \bv^T] \\
    &=  \bv (\bV^T \bV)^{-1} \bV^T \text{E}_{\boldsymbol{\Psi} \vert \boldsymbol{D}} [\bAone \bAone^T] \bV (\bV^T \bV)^{-1} \bv^T \\
    & \leq \lambda_{A_{11}} \bv (\bV^T \bV)^{-1} (\bV^T \bV)^{-1} \bv^T \\
\end{align*}
where $\lambda_{A_{11}}$ is the largest eigenvalue of $\bV^T \text{E}_{\boldsymbol{\Psi} \vert \boldsymbol{D}} [\bAone \bAone^T] \bV$. We need this quantity to be $o_p(n^{-1})$, which amounts to showing that $\lambda_{A_{11}} = o_p(n)$, since $\bv (\bV^T \bV)^{-1} (\bV^T \bV)^{-1} \bv^T = O_p(n^{-2})$. We know that the largest eigenvalue of a matrix is bounded above by the largest row sum of the matrix. The matrix $\bV^T \text{E}_{\boldsymbol{\Psi} \vert \boldsymbol{D}} [\bAone \bAone^T] \bV$ is a $q \times q$ matrix, where $q$ is finite-dimensional and therefore it suffices to show that any element of this matrix is $o_p(n)$. By simply performing matrix multiplication, one can show that the $(i,j)$ element of $\bV^T \text{E}_{\boldsymbol{\Psi} \vert \boldsymbol{D}} [\bAone \bAone^T] \bV$ is given by $$\sum_{a=1}^n V_{aj} \sum_{b=1}^n V_{bi} \text{E}_{\boldsymbol{\Psi} \vert \boldsymbol{D}} \Bigg[ (m_{ta} - m_{ta}^*) \Bigg(1 - \frac{1(T_a = t)}{p_{ta}^*} \Bigg) (m_{tb} - m_{tb}^*) \Bigg(1 - \frac{1(T_b = t)}{p_{tb}^*} \Bigg) \Bigg].$$
Now it suffices to show that this quantity is $o_p(n)$. To do so, we will show that for any $a$, we have that 
$$\sum_{b=1}^n V_{bi} \text{E}_{\boldsymbol{\Psi} \vert \boldsymbol{D}} \Bigg[ (m_{ta} - m_{ta}^*) \Bigg(1 - \frac{1(T_a = t)}{p_{ta}^*} \Bigg) (m_{tb} - m_{tb}^*) \Bigg(1 - \frac{1(T_b = t)}{p_{tb}^*} \Bigg) \Bigg] = o_p(1),$$
since a summation of $n$ of these terms would be guaranteed to be $o_p(n)$. We can re-write this expression as follows: 
$$\sum_{b=1}^n V_{bi} \Bigg(1 - \frac{1(T_a = t)}{p_{ta}^*} \Bigg) \Bigg(1 - \frac{1(T_b = t)}{p_{tb}^*} \Bigg) \text{E}_{\boldsymbol{\Psi} \vert \boldsymbol{D}} \Bigg[ (m_{ta} - m_{ta}^*)  (m_{tb} - m_{tb}^*) \Bigg].$$
In order to show that this term is $o_p(1)$, we will use sample splitting so that half of the data is used to find the posterior distribution of $\boldsymbol{\Psi} \vert \boldsymbol{D}$ while the remaining half is used to estimate the heterogeneous treatment effect function. When we refer to $n$ here, we are referring to the total sample size used in the second split, which is not used to estimate the posterior distribution. 
Due to the sample splitting, each of these $n$ terms is independent with variance given by 
$$\text{Var}_{\bD} \Bigg\{ V_{bi} \Bigg(1 - \frac{1(T_a = t)}{p_{ta}^*} \Bigg) \Bigg(1 - \frac{1(T_b = t)}{p_{tb}^*} \Bigg) \text{E}_{\boldsymbol{\Psi} \vert \boldsymbol{D}} \Bigg[ (m_{ta} - m_{ta}^*)  (m_{tb} - m_{tb}^*) \Bigg] \Bigg\}.$$ 
This variance can be bounded above by $$C_{A_{11}}  \text{Var}_{\bD} \Bigg\{ \text{E}_{\boldsymbol{\Psi} \vert \boldsymbol{D}} \Bigg[ (m_{ta} - m_{ta}^*)  (m_{tb} - m_{tb}^*) \Bigg] \Bigg\},$$
where $C_{A_{11}}$ is some fixed constant, due to positivity and the bounded covariates condition. Additionally, it is straightforward to show that each of these $n$ terms has expectation $0$ with respect to $\bD$ and therefore we can use the Lyapunov central limit theorem to state that 
$$\frac{1}{s_n} \sum_{b=1}^n V_{bi} \Bigg(1 - \frac{1(T_a = t)}{p_{ta}^*} \Bigg) \Bigg(1 - \frac{1(T_b = t)}{p_{tb}^*} \Bigg) \text{E}_{\boldsymbol{\Psi} \vert \boldsymbol{D}} \Bigg[ (m_{ta} - m_{ta}^*)  (m_{tb} - m_{tb}^*) \Bigg] = O_p(1),$$
where 
\begin{align*}
    s_n &= \sqrt{\sum_{b=1}^n \text{Var}_{\bD} \Bigg\{ V_{bi} \Bigg(1 - \frac{1(T_a = t)}{p_{ta}^*} \Bigg) \Bigg(1 - \frac{1(T_b = t)}{p_{tb}^*} \Bigg) \text{E}_{\boldsymbol{\Psi} \vert \boldsymbol{D}} \Bigg[ (m_{ta} - m_{ta}^*)  (m_{tb} - m_{tb}^*) \Bigg] \Bigg\}} \\
    & \leq C_{A_{12}} \sqrt{\sum_{b=1}^n   \text{Var}_{\bD} \Bigg\{ \text{E}_{\boldsymbol{\Psi} \vert \boldsymbol{D}} \Bigg[ (m_{ta} - m_{ta}^*)  (m_{tb} - m_{tb}^*) \Bigg] \Bigg\}} \\
    & \leq C_{A_{12}} \sqrt{\sum_{b=1}^n   \text{E}_{\bD} \Bigg\{ \text{E}_{\boldsymbol{\Psi} \vert \boldsymbol{D}} \Bigg[ (m_{ta} - m_{ta}^*)  (m_{tb} - m_{tb}^*) \Bigg]^2 \Bigg\}} \\
    &= o(1)
\end{align*}
where the last equality held by our moment assumption in Assumption \ref{asmpt: moments}. 
This, combined with the central limit theorem result above, gives us the desired result that 
\begin{align*}
    & \sum_{b=1}^n V_{bi} \Bigg(1 - \frac{1(T_a = t)}{p_{ta}^*} \Bigg) \Bigg(1 - \frac{1(T_b = t)}{p_{tb}^*} \Bigg) \text{E}_{\boldsymbol{\Psi} \vert \boldsymbol{D}} \Bigg[ (m_{ta} - m_{ta}^*)  (m_{tb} - m_{tb}^*) \Bigg] \\
    = & s_n \Bigg\{ \frac{1}{s_n} \sum_{b=1}^n V_{bi} \Bigg(1 - \frac{1(T_a = t)}{p_{ta}^*} \Bigg) \Bigg(1 - \frac{1(T_b = t)}{p_{tb}^*} \Bigg) \text{E}_{\boldsymbol{\Psi} \vert \boldsymbol{D}} \Bigg[ (m_{ta} - m_{ta}^*)  (m_{tb} - m_{tb}^*) \Bigg] \Bigg\}  \\
    = & o(1) O_p(1) = o_p(1)
\end{align*}
$\bAtwo$: The proof for $\bAtwo$ is analogous to the proof for $\bAone$ so it is left out for brevity. \\
$\bAthree$: 
\begin{align*}
    \text{Var}_{\boldsymbol{\Psi} \vert \boldsymbol{D}} [\bv (\bV^T \bV)^{-1} \bV^T \bAthree] &\leq \text{E}_{\boldsymbol{\Psi} \vert \boldsymbol{D}} [\bv (\bV^T \bV)^{-1} \bV^T \bAthree \bAthree^T \bV (\bV^T \bV)^{-1} \bv^T] \\
    &=  \bv (\bV^T \bV)^{-1} \bV^T \text{E}_{\boldsymbol{\Psi} \vert \boldsymbol{D}} [\bAthree \bAthree^T] \bV (\bV^T \bV)^{-1} \bv^T \\
    & \leq \lambda_{A_{31}} \bv (\bV^T \bV)^{-1} (\bV^T \bV)^{-1} \bv^T \\
\end{align*}
where $\lambda_{A_{31}}$ is the largest eigenvalue of $\bV^T \text{E}_{\boldsymbol{\Psi} \vert \boldsymbol{D}} [\bAthree \bAthree^T] \bV$. We need this quantity to be $o_p(n^{-1})$, which amounts to showing that $\lambda_{A_{31}} = o_p(n)$, since $\bv (\bV^T \bV)^{-1} (\bV^T \bV)^{-1} \bv^T = O_p(n^{-2})$. We know that the largest eigenvalue of a matrix is bounded above by the largest row sum of the matrix. The matrix $\bV^T \text{E}_{\boldsymbol{\Psi} \vert \boldsymbol{D}} [\bAthree \bAthree^T] \bV$ is a $q \times q$ matrix, where $q$ is finite-dimensional and therefore it suffices to show that any element of this matrix is $o_p(n)$. By simply performing matrix multiplication, one can show that the $(i,j)$ element of $\bV^T \text{E}_{\boldsymbol{\Psi} \vert \boldsymbol{D}} [\bAthree \bAthree^T] \bV$ is given by $$\sum_{a=1}^n V_{aj} \sum_{b=1}^n V_{bi} \text{E}_{\boldsymbol{\Psi} \vert \boldsymbol{D}} \bigg[ \frac{1(T_a = t)(p_{ta} - p_{ta}^*) (m_{ta} - m_{ta}^*) 1(T_b = t)(p_{tb} - p_{tb}^*) (m_{tb} - m_{tb}^*)}{p_{ta}p_{ta}^* p_{tb}p_{tb}^*} \bigg].$$
Now it suffices to show that this quantity is $o_p(n)$:
\begin{align*}
    & \sum_{a=1}^n V_{aj} \sum_{b=1}^n V_{bi} \text{E}_{\boldsymbol{\Psi} \vert \boldsymbol{D}} \bigg[ \frac{1(T_a = t)(p_{ta} - p_{ta}^*) (m_{ta} - m_{ta}^*) 1(T_b = t)(p_{tb} - p_{tb}^*) (m_{tb} - m_{tb}^*)}{p_{ta}p_{ta}^* p_{tb}p_{tb}^*} \bigg] \\
    & \leq \ \sum_{a=1}^n \sum_{b=1}^n \Bigg| V_{aj} V_{bi} \Bigg| \Bigg| \text{E}_{\boldsymbol{\Psi} \vert \boldsymbol{D}} \bigg[ \frac{1(T_a = t)(p_{ta} - p_{ta}^*) (m_{ta} - m_{ta}^*) 1(T_b = t)(p_{tb} - p_{tb}^*) (m_{tb} - m_{tb}^*)}{p_{ta}p_{ta}^* p_{tb}p_{tb}^*} \bigg] \Bigg| \\
    & \leq \ k \sum_{a=1}^n \sum_{b=1}^n \Bigg| \text{E}_{\boldsymbol{\Psi} \vert \boldsymbol{D}} \bigg[ \frac{1(T_a = t)(p_{ta} - p_{ta}^*) (m_{ta} - m_{ta}^*) 1(T_b = t)(p_{tb} - p_{tb}^*) (m_{tb} - m_{tb}^*)}{p_{ta}p_{ta}^* p_{tb}p_{tb}^*} \bigg] \Bigg| \\
    & \leq \ k \sum_{a=1}^n \sum_{b=1}^n \sqrt{\text{E}_{\boldsymbol{\Psi} \vert \boldsymbol{D}} \bigg[(m_{ta} - m_{ta}^*)^2 (m_{tb} - m_{tb}^*)^2 \bigg]} \\
    & \hspace{1.5 in}\times\sqrt{\text{E}_{\boldsymbol{\Psi} \vert \boldsymbol{D}} \bigg[ \frac{1(T_a = t)(p_{ta} - p_{ta}^*)^2 1(T_b = t)(p_{tb} - p_{tb}^*)^2}{(p_{ta}p_{ta}^* p_{tb}p_{tb}^*)^2} \bigg]} \\
    & \leq \ k' \sum_{a=1}^n \sum_{b=1}^n \sqrt{\text{E}_{\boldsymbol{\Psi} \vert \boldsymbol{D}} \bigg[(m_{ta} - m_{ta}^*)^2 (m_{tb} - m_{tb}^*)^2 \bigg]} \sqrt{\text{E}_{\boldsymbol{\Psi} \vert \boldsymbol{D}} \bigg[(p_{ta} - p_{ta}^*)^2 (p_{tb} - p_{tb}^*)^2 \bigg]} \\
    & = \ o_p(n) 
\end{align*}

The second inequality held because the covariates are bounded, while the fourth inequality held because of positivity and Assumption \ref{asmpt: posterior bounds}. The final assumption comes from our rate assumptions as defined in Assumption \ref{asmpt: moments} that describe how quickly certain moments of the posterior distribution must converge to zero. This concludes the proof of the fact that $\text{Var}_{\boldsymbol{\Psi} \vert \boldsymbol{D}} [\Delta(\boldsymbol{D}, \boldsymbol{\Psi})] = o_p(n^{-1})$.

\subsubsection{Asymptotic Behavior of $\text{Var}_{\boldsymbol{D}} \{ E_{\boldsymbol{\Psi} \vert \boldsymbol{D}} [\Delta(\boldsymbol{D}, \boldsymbol{\Psi})] \}$}
\label{sec:behavior of the true variance}

Now we will examine the behavior of the true variance of our estimator. Again we will do this for each of the four separate components separately, starting with $B$, which is straightforward because $\text{Var}_{\boldsymbol{D}} \{ E_{\boldsymbol{\Psi} \vert \boldsymbol{D}} [\bv (\bV^T \bV)^{-1} \bV^T \bB] \} = \text{Var}_{\boldsymbol{D}}[\bv (\bV^T \bV)^{-1} \bV^T \bB]$. We will now show that each of the remaining three terms are $o_p(n^{-1})$: \\
\\
$\bAone$: \begin{align*}
    \text{Var}_{\bD} \{ E_{\boldsymbol{\Psi} \vert \boldsymbol{D}} [\bv (\bV^T \bV)^{-1} \bV^T \bAone]\} 
    &= \text{Var}_{\bD} \{ [\bv (\bV^T \bV)^{-1} \bV^T E_{\boldsymbol{\Psi} \vert \boldsymbol{D}}[\bAone]\} \\
    &\leq \text{E}_{\bD} \{ \bv (\bV^T \bV)^{-1} \bV^T E_{\boldsymbol{\Psi} \vert \boldsymbol{D}}[\bAone] E_{\boldsymbol{\Psi} \vert \boldsymbol{D}}[\bAone]^T \bV (\bV^T \bV)^{-1} \bv^T \} \\
    & \leq \lambda_{A_{12}} \text{E}_{\bD} [\bv (\bV^T \bV)^{-1} (\bV^T \bV)^{-1} \bv^T] \\
\end{align*}
where $\lambda_{A_{12}}$ is the largest eigenvalue of $\bV^T E_{\boldsymbol{\Psi} \vert \boldsymbol{D}}[\bAone] E_{\boldsymbol{\Psi} \vert \boldsymbol{D}}[\bAone]^T \bV$. We need this quantity to be $o_p(n^{-1})$, which amounts to showing that $\lambda_{A_{12}} = o_p(n)$, since $\text{E}_{\bD}[\bv (\bV^T \bV)^{-1} (\bV^T \bV)^{-1} \bv^T] = O_p(n^{-2})$. We know that the largest eigenvalue of a matrix is bounded above by the largest row sum of the matrix. The matrix $\bV^T E_{\boldsymbol{\Psi} \vert \boldsymbol{D}}[\bAone] E_{\boldsymbol{\Psi} \vert \boldsymbol{D}}[\bAone]^T \bV$ is a $q \times q$ matrix, where $q$ is finite-dimensional and therefore it suffices to show that any element of this matrix is $o_p(n)$. One can show that the $(i,j)$ element of $\bV^T E_{\boldsymbol{\Psi} \vert \boldsymbol{D}}[\bAone] E_{\boldsymbol{\Psi} \vert \boldsymbol{D}}[\bAone]^T \bV$ is given by $$\sum_{a=1}^n V_{aj} \sum_{b=1}^n V_{bi} \text{E}_{\boldsymbol{\Psi} \vert \boldsymbol{D}} \Bigg[ (m_{ta} - m_{ta}^*) \Bigg(1 - \frac{1(T_a = t)}{p_{ta}^*} \Bigg) \Bigg] \text{E}_{\boldsymbol{\Psi} \vert \boldsymbol{D}} \Bigg[(m_{tb} - m_{tb}^*) \Bigg(1 - \frac{1(T_b = t)}{p_{tb}^*} \Bigg) \Bigg].$$
Now it suffices to show that this quantity is $o_p(n)$. To do so, we will show that for any $a$, we have that
$$\sum_{b=1}^n V_{bi} \text{E}_{\boldsymbol{\Psi} \vert \boldsymbol{D}} \Bigg[ (m_{ta} - m_{ta}^*) \Bigg(1 - \frac{1(T_a = t)}{p_{ta}^*} \Bigg) \Bigg] \text{E}_{\boldsymbol{\Psi} \vert \boldsymbol{D}} \Bigg[(m_{tb} - m_{tb}^*) \Bigg(1 - \frac{1(T_b = t)}{p_{tb}^*} \Bigg) \Bigg] = o_p(1),$$
since a summation of $n$ of these terms would be guaranteed to be $o_p(n)$. We can re-write this expression as follows: 
$$\sum_{b=1}^n V_{bi} \Bigg(1 - \frac{1(T_a = t)}{p_{ta}^*} \Bigg) \Bigg(1 - \frac{1(T_b = t)}{p_{tb}^*} \Bigg) \text{E}_{\boldsymbol{\Psi} \vert \boldsymbol{D}} \Bigg[ (m_{ta} - m_{ta}^*) \Bigg]  \text{E}_{\boldsymbol{\Psi} \vert \boldsymbol{D}} \Bigg[(m_{tb} - m_{tb}^*) \Bigg].$$
In order to show that this term is $o_p(1)$, we will use sample splitting so that half of the data is used to find the posterior distribution of $\boldsymbol{\Psi} \vert \boldsymbol{D}$ while the remaining half is used to estimate the heterogeneous treatment effect function. When we refer to $n$ here, we are referring to the total sample size used in the second split, which is not used to estimate the posterior distribution. Due to the sample splitting, each of these $n$ terms is independent with variance given by 
$$\text{Var}_{\bD} \Bigg\{ V_{bi} \Bigg(1 - \frac{1(T_a = t)}{p_{ta}^*} \Bigg) \Bigg(1 - \frac{1(T_b = t)}{p_{tb}^*} \Bigg) \text{E}_{\boldsymbol{\Psi} \vert \boldsymbol{D}} \Bigg[ (m_{ta} - m_{ta}^*) \Bigg]  \text{E}_{\boldsymbol{\Psi} \vert \boldsymbol{D}} \Bigg[ (m_{tb} - m_{tb}^*) \Bigg] \Bigg\}.$$ 
This variance can be bounded above by $$C_{A_{13}}  \text{Var}_{\bD} \Bigg\{ \text{E}_{\boldsymbol{\Psi} \vert \boldsymbol{D}} \Bigg[ (m_{ta} - m_{ta}^*) \Bigg]  \text{E}_{\boldsymbol{\Psi} \vert \boldsymbol{D}} \Bigg[ (m_{tb} - m_{tb}^*) \Bigg] \Bigg\},$$
where $C_{A_{13}}$ is some fixed constant, due to positivity and the bounded covariates condition. Additionally, it is straightforward to show that each of these $n$ terms has expectation $0$ with respect to $\bD$ and therefore we can use the Lyapunov central limit theorem to state that 
$$\frac{1}{s_n} \sum_{b=1}^n V_{bi} \Bigg(1 - \frac{1(T_a = t)}{p_{ta}^*} \Bigg) \Bigg(1 - \frac{1(T_b = t)}{p_{tb}^*} \Bigg) \text{E}_{\boldsymbol{\Psi} \vert \boldsymbol{D}} \Bigg[ (m_{ta} - m_{ta}^*) \Bigg]  \text{E}_{\boldsymbol{\Psi} \vert \boldsymbol{D}} \Bigg[ (m_{tb} - m_{tb}^*) \Bigg] = O_p(1),$$
where 
\begin{align*}
    s_n &= \sqrt{
    \splitfrac{\sum_{b=1}^n \text{Var}_{\bD} \Bigg\{ V_{bi} \Bigg(1 - \dfrac{1(T_a = t)}{p_{ta}^*} \Bigg) \Bigg(1 - \dfrac{1(T_b = t)}{p_{tb}^*} \Bigg)}{\hspace{1 in}\times\text{E}_{\boldsymbol{\Psi} \vert \boldsymbol{D}} \Bigg[ (m_{ta} - m_{ta}^*)\Bigg]  \text{E}_{\boldsymbol{\Psi} \vert \boldsymbol{D}} \Bigg[ (m_{tb} - m_{tb}^*) \Bigg] \Bigg\}}
    } \\
    & \leq C_{A_{14}} \sqrt{\sum_{b=1}^n   \text{Var}_{\bD} \Bigg\{ \text{E}_{\boldsymbol{\Psi} \vert \boldsymbol{D}} \Bigg[ (m_{ta} - m_{ta}^*)  \Bigg]  \text{E}_{\boldsymbol{\Psi} \vert \boldsymbol{D}} \Bigg[ (m_{tb} - m_{tb}^*) \Bigg] \Bigg\}} \\
    & \leq C_{A_{14}} \sqrt{\sum_{b=1}^n   \text{E}_{\bD} \Bigg\{ \text{E}_{\boldsymbol{\Psi} \vert \boldsymbol{D}} \Bigg[ (m_{ta} - m_{ta}^*) \Bigg]^2  \text{E}_{\boldsymbol{\Psi} \vert \boldsymbol{D}} \Bigg[ (m_{tb} - m_{tb}^*) \Bigg]^2 \Bigg\}} \\
    &= o(1)
\end{align*}
where the last equality held by our moment assumption in Assumption \ref{asmpt: moments}. This, combined with the central limit theorem result above, gives us the desired result that 
\begin{align*}
    & \sum_{b=1}^n V_{bi} \Bigg(1 - \frac{1(T_a = t)}{p_{ta}^*} \Bigg) \Bigg(1 - \frac{1(T_b = t)}{p_{tb}^*} \Bigg) \text{E}_{\boldsymbol{\Psi} \vert \boldsymbol{D}} \Bigg[ (m_{ta} - m_{ta}^*) \Bigg]  \text{E}_{\boldsymbol{\Psi} \vert \boldsymbol{D}} \Bigg[ (m_{tb} - m_{tb}^*) \Bigg] \\
    = & s_n \Bigg\{ \frac{1}{s_n} \sum_{b=1}^n V_{bi} \Bigg(1 - \frac{1(T_a = t)}{p_{ta}^*} \Bigg) \Bigg(1 - \frac{1(T_b = t)}{p_{tb}^*} \Bigg) \text{E}_{\boldsymbol{\Psi} \vert \boldsymbol{D}} \Bigg[ (m_{ta} - m_{ta}^*) \Bigg]  \text{E}_{\boldsymbol{\Psi} \vert \boldsymbol{D}} \Bigg[ (m_{tb} - m_{tb}^*) \Bigg] \Bigg\}  \\
    = & o(1) O_p(1) = o_p(1)
\end{align*}
$\bAtwo$: The proof for $\bAtwo$ is analogous to the proof for $\bAone$ so it is left out for brevity. \\
$\bAthree:$ \begin{align*}
    \text{Var}_{\bD} \{ E_{\boldsymbol{\Psi} \vert \boldsymbol{D}} [\bv (\bV^T \bV)^{-1} \bV^T \bAthree] \} 
    &= \text{Var}_{\bD} \{ \bv (\bV^T \bV)^{-1} \bV^T E_{\boldsymbol{\Psi} \vert \boldsymbol{D}} [\bAthree] \} \\
    &\leq \text{E}_{\bD} \{ \bv (\bV^T \bV)^{-1} \bV^T \text{E}_{\boldsymbol{\Psi} \vert \boldsymbol{D}}[\bAthree] \text{E}_{\boldsymbol{\Psi} \vert \boldsymbol{D}}[\bAthree]^T \bV (\bV^T \bV)^{-1} \bv^T \} \\
    & \leq \lambda_{A_{32}} \text{E}_{\bD} [\bv (\bV^T \bV)^{-1} (\bV^T \bV)^{-1} \bv^T] \\
\end{align*}
where $\lambda_{A_{32}}$ is the largest eigenvalue of $\bV^T \text{E}_{\boldsymbol{\Psi} \vert \boldsymbol{D}} [\bAthree] \text{E}_{\boldsymbol{\Psi} \vert \boldsymbol{D}} [\bAthree]^T \bV$. We need this quantity to be $o_p(n^{-1})$, which amounts to showing that $\lambda_{A_{32}} = o_p(n)$, since $\text{E}_{\bD} [\bv (\bV^T \bV)^{-1} (\bV^T \bV)^{-1} \bv^T] = O_p(n^{-2})$. We know that the largest eigenvalue of a matrix is bounded above by the largest row sum of the matrix. The matrix $\bV^T \text{E}_{\boldsymbol{\Psi} \vert \boldsymbol{D}} [\bAthree] \text{E}_{\boldsymbol{\Psi} \vert \boldsymbol{D}} [\bAthree]^T \bV$ is a $q \times q$ matrix, where $q$ is finite-dimensional and therefore it suffices to show that any element of this matrix is $o_p(n)$. One can show that the $(i,j)$ element of $\bV^T \text{E}_{\boldsymbol{\Psi} \vert \boldsymbol{D}} [\bAthree] \text{E}_{\boldsymbol{\Psi} \vert \boldsymbol{D}} [\bAthree]^T \bV$ is given by $$\sum_{a=1}^n V_{aj} \sum_{b=1}^n V_{bi} \text{E}_{\boldsymbol{\Psi} \vert \boldsymbol{D}} \bigg[ \frac{1(T_a = t)(p_{ta} - p_{ta}^*) (m_{ta} - m_{ta}^*)}{p_{ta}p_{ta}^*} \bigg] \text{E}_{\boldsymbol{\Psi} \vert \boldsymbol{D}} \bigg[ \frac{1(T_b = t)(p_{tb} - p_{tb}^*) (m_{tb} - m_{tb}^*)}{p_{tb}p_{tb}^*} \bigg].$$
Now it suffices to show that this quantity is $o_p(n)$:
\begin{align*}
    & \sum_{a=1}^n V_{aj} \sum_{b=1}^n V_{bi} \text{E}_{\boldsymbol{\Psi} \vert \boldsymbol{D}} \bigg[ \frac{1(T_a = t)(p_{ta} - p_{ta}^*) (m_{ta} - m_{ta}^*)}{p_{ta}p_{ta}^*} \bigg] \text{E}_{\boldsymbol{\Psi} \vert \boldsymbol{D}} \bigg[ \frac{1(T_b = t)(p_{tb} - p_{tb}^*) (m_{tb} - m_{tb}^*)}{p_{tb}p_{tb}^*} \bigg] \\
    & \leq \ \sum_{a=1}^n \sum_{b=1}^n \Bigg| V_{aj} V_{bi} \Bigg| \Bigg| \text{E}_{\boldsymbol{\Psi} \vert \boldsymbol{D}} \bigg[ \frac{1(T_a = t)(p_{ta} - p_{ta}^*) (m_{ta} - m_{ta}^*)}{p_{ta}p_{ta}^*} \bigg] \Bigg|\\
    &\hspace{2.5 in}\times\Bigg| \text{E}_{\boldsymbol{\Psi} \vert \boldsymbol{D}} \bigg[ \frac{1(T_b = t)(p_{tb} - p_{tb}^*) (m_{tb} - m_{tb}^*)}{p_{tb}p_{tb}^*} \bigg] \Bigg| \\
    & \leq \ k'' \sum_{a=1}^n \sum_{b=1}^n \Bigg| \text{E}_{\boldsymbol{\Psi} \vert \boldsymbol{D}} \bigg[ \frac{1(T_a = t)(p_{ta} - p_{ta}^*) (m_{ta} - m_{ta}^*)}{p_{ta}p_{ta}^*} \bigg] \Bigg|\\
    &\hspace{2.5 in}\times \Bigg| \text{E}_{\boldsymbol{\Psi} \vert \boldsymbol{D}} \bigg[ \frac{1(T_b = t)(p_{tb} - p_{tb}^*) (m_{tb} - m_{tb}^*)}{p_{tb}p_{tb}^*} \bigg] \Bigg| \\
    & \leq \ k'' \sum_{a=1}^n \sum_{b=1}^n \sqrt{\text{E}_{\boldsymbol{\Psi} \vert \boldsymbol{D}} \bigg[(m_{ta} - m_{ta}^*)^2 \bigg] \text{E}_{\boldsymbol{\Psi} \vert \boldsymbol{D}} \bigg[\frac{1(T_a = t)(p_{ta} - p_{ta}^*)^2}{(p_{ta} p_{ta}^*)^2} \bigg]}  \\
    &\hspace{2.5 in} \times \sqrt{\text{E}_{\boldsymbol{\Psi} \vert \boldsymbol{D}} \bigg[(m_{tb} - m_{tb}^*)^2 \bigg] \text{E}_{\boldsymbol{\Psi} \vert \boldsymbol{D}} \bigg[\frac{1(T_b = t)(p_{tb} - p_{tb}^*)^2}{(p_{tb} p_{tb}^*)^2} \bigg]} \\
    & \leq \ k''' \sum_{a=1}^n \sum_{b=1}^n \sqrt{\text{E}_{\boldsymbol{\Psi} \vert \boldsymbol{D}} \bigg[(m_{ta} - m_{ta}^*)^2 \bigg] \text{E}_{\boldsymbol{\Psi} \vert \boldsymbol{D}} \bigg[(p_{ta} - p_{ta}^*)^2 \bigg]} \\
    &\hspace{2.5 in}\times \sqrt{\text{E}_{\boldsymbol{\Psi} \vert \boldsymbol{D}} \bigg[(m_{tb} - m_{tb}^*)^2 \bigg] \text{E}_{\boldsymbol{\Psi} \vert \boldsymbol{D}} \bigg[(p_{tb} - p_{tb}^*)^2 \bigg]} \\
    & = \ o_p(n) 
\end{align*}

The second inequality held because the covariates are bounded, while the fourth inequality held because of positivity and Assumption \ref{asmpt: posterior bounds}. The final equality comes from our rate assumptions as defined in Assumption \ref{asmpt: moments}. This concludes the proof of the fact that $\text{Var}_{\boldsymbol{D}} \{ E_{\boldsymbol{\Psi} \vert \boldsymbol{D}} [\Delta(\boldsymbol{D}, \boldsymbol{\Psi})] \} = \text{Var}_{\boldsymbol{D}}\Big[ \bv (\bV^T \bV)^{-1} \bV^T \bB \Big] + o_p(n^{-1})$. \\

\subsubsection{Asymptotic Behavior of $\text{Var}_{\boldsymbol{D^{(m)}}} \{ E_{\boldsymbol{\Psi} \vert \boldsymbol{D}} [\Delta(\boldsymbol{D^{(m)}}, \boldsymbol{\Psi})] \}$}

We will again begin by detailing the asymptotic behavior of each of the four components of $\Delta(\boldsymbol{D}, \boldsymbol{\Psi})$ separately. We will first examine the $\bB$ component. To do this, define $\bv_m$, $\bV_m$, and $\bB_m$ to be length $n$ vectors of resampled values from the $\bv$, $\bV$, and $\bB$ vectors, respectively. 
\begin{align*}
    \text{Var}_{\boldsymbol{D^{(m)}}} \{ E_{\boldsymbol{\Psi} \vert \boldsymbol{D}} [\bv_m (\bV_m^T \bV_m)^{-1} \bV_m^T \bB_m] \} &= \text{Var}_{\boldsymbol{D^{(m)}}} [\bv_m (\bV_m^T \bV_m)^{-1} \bV_m^T \bB_m] \\
    &\approx \text{Var}_{\boldsymbol{D}} [\bv (\bV^T \bV)^{-1} \bV^T \bB]
\end{align*}
The approximation stems from the fact that we are using the bootstrap to approximate the distribution of $\boldsymbol{D}$, which is justified asymptotically. Effectively, we are using the bootstrap here to approximate the variance of a linear regression prediction, which should hold well in practice. What remains to show is that the remaining terms involving $\bAone$, $\bAtwo$, and $\bAthree$ are all $o_p(n^{-1})$. The proofs for each of these three terms are identical to the corresponding proofs from the previous section involving $\text{Var}_{\boldsymbol{D}} \{ E_{\boldsymbol{\Psi} \vert \boldsymbol{D}} [\Delta(\boldsymbol{D}, \boldsymbol{\Psi})] \}$ with the outer moment changed to be with respect to $\boldsymbol{D^{(m)}}$ instead of $\boldsymbol{D}$, and all other ideas follow analogously.

\section{Conservativeness of Variance Procedure}\label{sec:conservativeness}
Here, we show that our variance estimator tends to be conservative in the sense that it is larger than the true variance on average. First, we split the data $\bD$ into $\bD_{(1)}$ and $\bD_{(2)}$, used to find the posterior distribution and estimate the heterogeneous treatment effect function, respectively. As a reminder, under sample splitting, we can denote the true variance by 
\begin{align*}
    V&=\text{Var}_{\{\bD_{(1)},\bD_{(2)}\}} \left\{E_{\bPsi | \bD_{(1)}}[\Delta(\bPsi, \bD_{(2)})]\right\}\\
    &=E_{\bD_{(1)}}\left[\text{Var}_{\bD_{(2)}|\bD_{(1)}} \{ E_{\bPsi | \bD_{(1)}}[\Delta(\bPsi, \bD_{(2)})] \}\right] +\text{Var}_{\bD_{(1)}}\left[E_{\bD_{(2)}|\bD_{(1)}} \{ E_{\bPsi | \bD_{(1)}}[\Delta(\bPsi, \bD_{(2)})] \}\right] \\
    &=E_{\bD_{(1)}}\left[\text{Var}_{\bD_{(2)}} \{ E_{\bPsi | \bD_{(1)}}[\Delta(\bPsi, \bD_{(2)})] \}\right] +\text{Var}_{\bD_{(1)}}\left[E_{\bD_{(2)}} \{ E_{\bPsi | \bD_{(1)}}[\Delta(\bPsi, \bD_{(2)})] \} \right],
\end{align*}
where the third equality follows from the independence of $\bD_{(1)}$ and $\bD_{(2)}$. The variance estimator is given by
\begin{align*}
    \widehat{V}
    &= \text{Var}_{\bD^{(m)}_{(2)}} \{ E_{\bPsi | \bD_{(1)}}[\Delta(\bPsi, \bD^{(m)}_{(2)})] \} + \text{Var}_{\bPsi | \bD_{(1)}}[\Delta(\bPsi, \bD_{(2)})]\\
    &\approx \text{Var}_{\bD_{(2)}} \{ E_{\bPsi | \bD_{(1)}}[\Delta(\bPsi, \bD_{(2)})] \} + \text{Var}_{\bPsi | \bD_{(1)}}[\Delta(\bPsi, \bD_{(2)})]\\
    &\approx \text{Var}_{\bD_{(2)}} \{ E_{\bPsi | \bD_{(1)}}[\Delta(\bPsi, \bD_{(2)})] \} + \text{Var}_{\bD_{(1)}}\{ E_{\bPsi | \bD_{(1)}}[\Delta(\bPsi, \bD_{(2)})] \},
\end{align*}
where the first approximation stems from using the bootstrap variance to approximate the population variance. The second approximation should hold when a Berstein-von Mises theorem exists for the posterior distribution of the propensity score and outcome regression models. These are the basis for using posterior intervals for frequentist inference, and imply that the posterior variance should well approximate the sampling distribution variance of the posterior mean. Then, we have
\begin{align*}
    &E_{\{\bD_{(1)},\bD_{(2)}\}}(\widehat{V}-V)\\
    &\approx E_{\bD_{(1)}}\left[\text{Var}_{\bD_{(2)}} \{ E_{\bPsi | \bD_{(1)}}[\Delta(\bPsi, \bD_{(2)})] \}\right] + E_{\bD_{(2)}}\left[\text{Var}_{\bD_{(1)}}\{ E_{\bPsi | \bD_{(1)}}[\Delta(\bPsi, \bD_{(2)})] \} \right]\\
    &\hspace{.2in}-E_{\bD_{(1)}}\left[\text{Var}_{\bD_{(2)}} \{ E_{\bPsi | \bD_{(1)}}[\Delta(\bPsi, \bD_{(2)})] \}\right] -\text{Var}_{\bD_{(1)}}\left[E_{\bD_{(2)}} \{ E_{\bPsi | \bD_{(1)}}[\Delta(\bPsi, \bD_{(2)})] \}\right]\\
    &=E_{\bD_{(2)}}\left[\text{Var}_{\bD_{(1)}}\{ E_{\bPsi | \bD_{(1)}}[\Delta(\bPsi, \bD_{(2)})] \}\right]-\text{Var}_{\bD_{(1)}}\left[E_{\bD_{(2)}} \{ E_{\bPsi | \bD_{(1)}}[\Delta(\bPsi, \bD_{(2)})] \}\right]
\end{align*}
Now, it suffices to show the following inequality holds:
\begin{equation}\label{eq:conservative}
    E_{\bD_{(2)}}\left[\text{Var}_{\bD_{(1)}}\{ E_{\bPsi | \bD_{(1)}}[\Delta(\bPsi, \bD_{(2)})] \}\right]\geq \text{Var}_{\bD_{(1)}}\left[E_{\bD_{(2)}} \{ E_{\bPsi | \bD_{(1)}}[\Delta(\bPsi, \bD_{(2)})] \}\right].
\end{equation}
Observing that the posterior mean $E_{\bPsi | \bD_{(1)}}[\Delta(\bPsi, \bD_{(2)})]$ is a random variable itself, the above inequality means that the variance of the average of a random variable is less than the average of the variance of the random variable. Intuitively, we expect this to generally hold as first taking the average of the random variable can remove certain sources of variability. For the doubly robust estimator that we use in the manuscript, we can provide some additional intuition for why this is likely to hold.
Recall that our estimator of $E(Y(t)|\bV=\bv)$ is given by
$$\Delta(\bPsi, \bD_{(2)}) = \bv (\bV^T \bV)^{-1} \bV^T \bZ = \bv (\bV^T \bV)^{-1} \bV^T (\bAone + \bAtwo + \bAthree + \bB),$$
and note that we have
\begin{align*}
    E_{\bD_{(2)}} \{ E_{\bPsi | \bD_{(1)}}[\bv (\bV^T \bV)^{-1} \bV^T \bAone] \}
    &= E_{\bPsi | \bD_{(1)}}\{ E_{\bD_{(2)}}[\bv (\bV^T \bV)^{-1} \bV^T \bAone] \}\\
    &= E_{\bPsi | \bD_{(1)}}\{ E_{X}E_{T|X}E_{Y|X,T}[\bv (\bV^T \bV)^{-1} \bV^T \bAone] \} \\
    &= 0
\end{align*}
where the first equality follows from the independence of $\bD_{(1)}$ and $\bD_{(2)}$ and the last equality follows from
\begin{align*}
    E_{X}E_{T|X}E_{Y|X,T}[\lambda_iA_{1i}]
    &=E_{X}E_{T|X}E_{Y|X,T}\left[\lambda_i(m_{ti} - m_{ti}^*) \Bigg(1 - \frac{1(T_i = t)}{p_{ti}^*} \Bigg)\right]\\
    &=E_{X}E_{T|X}\left[\lambda_i(m_{ti} - m_{ti}^*) \Bigg(1 - \frac{1(T_i = t)}{p_{ti}^*} \Bigg)\right]\\
    &=E_{X}\left\{\lambda_i(m_{ti} - m_{ti}^*)E_{T|X}\left[ \Bigg(1 - \frac{1(T_i = t)}{p_{ti}^*} \Bigg)\right] \right\}\\
    &=E_{X}\left\{\lambda_i(m_{ti} - m_{ti}^*)\Bigg(1 - \frac{p_{ti}^*}{p_{ti}^*} \Bigg) \right\}\\
    &=0
\end{align*}
where $\lambda_i$ denotes the $i$-th element of $\bv(\bV^T \bV)^{-1} \bV^T$. Analogously, one can easily check that $E_{\bD_{(2)}} \{ E_{\bPsi | \bD_{(1)}}[\bv (\bV^T \bV)^{-1} \bV^T \bAtwo] \}=0$ and $E_{\bD_{(2)}} \{ E_{\bPsi | \bD_{(1)}}[\bv (\bV^T \bV)^{-1} \bV^T \bB] \}=E_{\bD_{(2)}} \{ E_{\bPsi | \bD_{(1)}}(\bB_2)\}$ where $\bB_2$ is a $n\times 1$ vector of $m_{ti}^*$. Hence, we can get rid of a number of terms by taking the expectation first and write
$$\text{Var}_{\bD_{(1)}}\left[E_{\bD_{(2)}} \{ E_{\bPsi | \bD_{(1)}}[\Delta(\bPsi,\bD_{(2)})] \}\right]=\text{Var}_{\bD_{(1)}}\left[E_{\bD_{(2)}} \{ E_{\bPsi | \bD_{(1)}}[\bv (\bV^T \bV)^{-1} \bV^T (\bAthree+\bB_2)] \}\right],$$
which effectively states that only $\bAthree$ and $\bB_2$ contribute to the overall variance, while all of the components of $\bZ$ contribute to the overall variance in $E_{\bD_{(2)}}\left[\text{Var}_{\bD_{(1)}}\{ E_{\bPsi | \bD_{(1)}}[\Delta(\bPsi, \bD_{(2)})] \}\right]$ since there is no such cancellation. Unless the terms in $\bAone$ and $\bAtwo$ are strongly, negatively correlated with those in $\bAthree$ and $\bB_2$, which we do not expect, these terms cancelling out should reduce the overall variability.  Therefore, we expect that  $$E_{\bD_{(2)}}\left[\text{Var}_{\bD_{(1)}}\{ E_{\bPsi | \bD_{(1)}}[\Delta(\bPsi, \bD_{(2)})] \}\right]\geq\text{Var}_{\bD_{(1)}}\left[E_{\bD_{(2)}} \{ E_{\bPsi | \bD_{(1)}}[\Delta(\bPsi, \bD_{(2)})] \}\right].$$
Now we will show that this can be rigorously proven for certain types of functions $\Delta(\bPsi, \bD_{(2)})$ and we will highlight how certain causal estimators fall within this scope.

\subsection{Conservativeness for Class of Functions}
For this section, let's assume that $\Delta(\bPsi, \bD_{(2)})=\boldsymbol g^T(\bD_{(2)}) \boldsymbol f(\bPsi)$ where $\boldsymbol f(\cdot)$ and $\boldsymbol g(\cdot)$ are some vector-valued functions. Then, we have
\begin{align*}
    E_{\bPsi | \bD_{(1)}}[\boldsymbol g^T(\bD_{(2)}) \boldsymbol f(\bPsi)]=\int \boldsymbol g^T(\bD_{(2)}) \boldsymbol f(\bPsi) dP(\bPsi | \bD_{(1)})
    &=\boldsymbol g^T(\bD_{(2)}) \int \boldsymbol f(\bPsi) dP(\bPsi | \bD_{(1)})\\
    &:=\boldsymbol g^T(\bD_{(2)}) \boldsymbol h(\bD_{(1)}),
\end{align*}
and thus, we have
\begin{align*}
    E_{\bD_{(2)}}\left[\text{Var}_{\bD_{(1)}}\{ \boldsymbol g^T(\bD_{(2)}) \boldsymbol h(\bD_{(1)}) \}\right]
    &= E_{\bD_{(2)}}\left[ \boldsymbol g^T(\bD_{(2)}) \text{Var}_{\bD_{(1)}}(\boldsymbol h(\bD_{(1)})) \boldsymbol g(\bD_{(2)})  \right]\\
    &= Tr(\text{Var}_{\bD_{(1)}}(\boldsymbol h(\bD_{(1)})) \boldsymbol{\Sigma}_G) \\
    &+ E_{\bD_{(2)}}\left[ \boldsymbol g^T(\bD_{(2)}) \right] \text{Var}_{\bD_{(1)}}(\boldsymbol h(\bD_{(1)})) E_{\bD_{(2)}}\left[ \boldsymbol g(\bD_{(2)}) \right]  \\
    &\geq E_{\bD_{(2)}}\left[ \boldsymbol g^T(\bD_{(2)}) \right] \text{Var}_{\bD_{(1)}}(\boldsymbol h(\bD_{(1)})) E_{\bD_{(2)}}\left[ \boldsymbol g(\bD_{(2)}) \right]\\
    \text{Var}_{\bD_{(1)}}\left[E_{\bD_{(2)}} \{ \boldsymbol g^T(\bD_{(2)}) \boldsymbol h(\bD_{(1)}) \}\right] 
    &= \text{Var}_{\bD_{(1)}}\left[\boldsymbol  E_{\bD_{(2)}} \{ \boldsymbol g^T(\bD_{(2)}) \} \boldsymbol h(\bD_{(1)}) \right]\\
    &=E_{\bD_{(2)}} [ \boldsymbol g^T(\bD_{(2)}) ] \text{Var}_{\bD_{(1)}}\left(\boldsymbol h(\bD_{(1)}) \right) E_{\bD_{(2)}} [ \boldsymbol g(\bD_{(2)}) ]
\end{align*}
where $\boldsymbol{\Sigma}_G$ denotes the variance-covariance matrix of $\boldsymbol g(\bD_{(2)})$. Therefore, Equation \ref{eq:conservative} holds.
We now give examples of estimators where $\Delta(\bPsi, \bD_{(2)})$ can be expressed as $\boldsymbol g^T(\bD_{(2)}) \boldsymbol f(\bPsi)$ and thus the variance estimators are conservative.

\begin{itemize}
    \item Outcome regression model: $\Delta(\bPsi, \bD_{(2)})=m_{ti}=\bv \boldsymbol\beta$ is a fitted value of the outcome regression model for $T=t$ with the regression coefficients vector $\boldsymbol{\beta}$. Note that this parameter vector does not depend on $\bD_{(2)}$. In this case, $\boldsymbol g^T(\bD_{(2)})=\bv$ and $\boldsymbol f(\bPsi)=\boldsymbol\beta$.
    
    
    \item IPW estimator with the misspecified propensity score: Let the propensity score be fixed over the covariates, say $\alpha$. Then,
    \begin{align*}
        \Delta(\bPsi, \bD_{(2)}) 
        &= \bv(\bV^T\bV)^{-1}\bV^T\boldsymbol{Z}\\
        &=\bv(\bV^T\bV)^{-1}\bV^T\boldsymbol{W}\dfrac{1}{\alpha}
    \end{align*}
    where $\boldsymbol{Z}=\dfrac{1}{\alpha}\boldsymbol{W}$ is a $n\times 1$ vector of $Z_{i}=\dfrac{1(T_i=t)Y_i}{p_{ti}}=\dfrac{1(T_i=t)Y_i}{\alpha}$ evaluated at each individual from $\bD_{(2)}$ and $\boldsymbol{W}$ is a vector of $1(T_i=t)Y_i$. In this case, $\boldsymbol g^T(\bD_{(2)})=\bv(\bV^T\bV)^{-1}\bV^T\boldsymbol{W}$ and $\boldsymbol f(\bPsi)=\dfrac{1}{\alpha}$.
    
    \item Doubly robust estimator with misspecified propensity score. Again, let the propensity score be fixed over the covariates, given by $\alpha$:
    \begin{align*}
        \Delta(\bPsi, \bD_{(2)}) &= \bv(\bV^T\bV)^{-1}\bV^T\boldsymbol{Z} = \bv(\bV^T\bV)^{-1}\bV^T \left(\dfrac{1}{\alpha}\boldsymbol{W}-\dfrac{\boldsymbol{\widetilde{X}\beta}}{\alpha}+\boldsymbol{X\beta}\right)\\
        &= \underbrace{\bv(\bV^T\bV)^{-1}\bV^T\boldsymbol{W}}_{\boldsymbol g_1^T(\bD_{(2)})}\underbrace{\dfrac{1}{\alpha}}_{\boldsymbol f_1(\bPsi)}\\
        &\hspace{1 in}-\underbrace{\bv(\bV^T\bV)^{-1}\bV^T\boldsymbol{\widetilde{X}}}_{\boldsymbol g_2^T(\bD_{(2)})}\underbrace{\dfrac{\boldsymbol{\beta}}{\alpha}}_{\boldsymbol f_2(\bPsi)}+\underbrace{\bv(\bV^T\bV)^{-1}\bV^T\boldsymbol X}_{\boldsymbol g_3^T(\bD_{(2)})}\underbrace{\boldsymbol{\beta}}_{\boldsymbol f_3^T(\bD_{(2)})}\\
        &= \boldsymbol g^T(\bD_{(2)}) \boldsymbol f(\bPsi)
    \end{align*}
    where $\boldsymbol{Z}$ is a $n\times 1$ vector of $Z_{i}=\dfrac{1(T_i = t)}{\alpha} (Y_i - m_{ti}) + m_{ti}$, $\boldsymbol{W}$ is a vector of $1(T_i=t)Y_i$, $\boldsymbol{\widetilde{X}}$ is identical to $\boldsymbol{X}$ but with the $i^{th}$ row equal to the zero vector when $T_i \neq t$, $\boldsymbol{g}^T=(\boldsymbol{g}^T_1,\boldsymbol{g}^T_2,\boldsymbol{g}^T_3)$, and $\boldsymbol{f}^T=(\boldsymbol{f}^T_1,\boldsymbol{f}^T_2,\boldsymbol{f}^T_3)$.
\end{itemize}

\subsection{An Upper Bound of the Variance estimator}\label{sec:upper bound}
We have shown that our variance estimator is conservative in that it is greater than or equal to the true variance on average. We will now prove that there is a reasonable upper bound of the proposed variance estimator so that it cannot be extremely high and yield meaningless inference. Recall that we can express the true variance as
$$V=E_{\bD_{(2)}}\left[\text{Var}_{\bD_{(1)}} \{ E_{\bPsi | \bD_{(1)}}[\Delta(\bPsi, \bD_{(2)})] \}\right] +\text{Var}_{\bD_{(2)}}\left[E_{\bD_{(1)}} \{ E_{\bPsi | \bD_{(1)}}[\Delta(\bPsi, \bD_{(2)})] \} \right],$$
and therefore,
\begin{align*}
    &E_{\{\bD_{(1)},\bD_{(2)}\}}(\widehat{V}-V)
    \\&\approx E_{\bD_{(2)}}\left[\text{Var}_{\bD_{(1)}}\{ E_{\bPsi | \bD_{(1)}}[\Delta(\bPsi, \bD_{(2)})] \}\right]-\text{Var}_{\bD_{(1)}}\left[E_{\bD_{(2)}} \{ E_{\bPsi | \bD_{(1)}}[\Delta(\bPsi, \bD_{(2)})] \}\right]\\
    &= V-\text{Var}_{\bD_{(1)}}\left[E_{\bD_{(2)}} \{ E_{\bPsi | \bD_{(1)}}[\Delta(\bPsi, \bD_{(2)})] \}\right]-\text{Var}_{\bD_{(2)}}\left[E_{\bD_{(1)}} \{ E_{\bPsi | \bD_{(1)}}[\Delta(\bPsi, \bD_{(2)})] \}\right]\\
    &\leq V - \max\left(\text{Var}_{\bD_{(1)}}\left[E_{\bD_{(2)}} \{ E_{\bPsi | \bD_{(1)}}[\Delta(\bPsi, \bD_{(2)})] \}\right],\text{Var}_{\bD_{(2)}}\left[E_{\bD_{(1)}} \{ E_{\bPsi | \bD_{(1)}}[\Delta(\bPsi, \bD_{(2)})] \}\right]\right)
\end{align*}
which illustrates that our variance estimator is substantially less than twice the true variance on average.

\section{Proof of consistency of the doubly robust estimator}
\label{sec:consistency of estimator}
In this section, we show that if one of the nuisance models is correctly specified, the posterior distribution of the doubly robust estimator contracts around the true CATE at a reasonable rate:
    $$\sup\limits_{P_0} E_{P_0} \mathbb{P}_n\left(|\Delta(\boldsymbol{D}, \boldsymbol{\Psi})-\tau(\bv)|>M\epsilon_n|\bD\right)\rightarrow0$$
    for some $\epsilon_n\rightarrow0$, and further our estimator, $E_{\bPsi|\bD}[\Delta(\boldsymbol{D}, \boldsymbol{\Psi})]$, is consistent for CATE assuming the bounded posterior variance of $\Delta(\boldsymbol{D}, \boldsymbol{\Psi})$.
    
We first decompose the pseudo-outcome $Z_{i}^{(t)}$ in a difference way,
    \begin{align*}
        Z_{i}^{(t)} &= \frac{1(T_i = t)}{p_{ti}} (Y_i - m_{ti}) + m_{ti}\\
        &= \frac{1(T_i = t)}{p_{ti}} (Y_i - m_{ti}) + m_{ti} \\
    &= (m_{ti} - \tilde{m}_{ti}) \Bigg(1 - \frac{1(T_i = t)}{\tilde{p}_{ti}} \Bigg) \\
    &+ \frac{1(T_i = t)(p_{ti} - \tilde{p}_{ti}) (\tilde{m}_{ti} - Y_i)}{p_{ti}\tilde{p}_{ti}} \\
    &+ \frac{1(T_i = t)(p_{ti} - \tilde{p}_{ti}) (m_{ti} - \tilde{m}_{ti})}{p_{ti}\tilde{p}_{ti}} \\
    &+ \frac{1(T_i = t)}{\tilde{p}_{ti}} (Y_i - \tilde{m}_{ti}) + \tilde{m}_{ti} \\
    &= A_{i1}^{(t)} + A_{i2}^{(t)} + A_{i3}^{(t)} + B_{i}^{(t)}.
    \end{align*}
    where $\tilde{p}_{ti}$ and $\tilde{m}_{ti}$ denote the limiting values of $p_{ti}$ and $m_{ti}$. Then, we define $Z_i\equiv Z_i^{(1)}-Z_i^{(0)}$ and $\boldsymbol{Z}$, a $n \times 1$ vector of this quantity evaluated at each of the $n$ data points. Similarly, we also define $\boldsymbol{A}_{1}$, $\boldsymbol{A}_{2}$, $\boldsymbol{A}_{3}$, and, $\boldsymbol{B}$. Given a parameter vector $\bPsi$, an estimator of $E(Y(1)-Y(0) | \bV = \bv)=\tau(\bv)$ can then be defined as
    \begin{align*}
        \Delta(\boldsymbol{D}, \boldsymbol{\Psi}) = \bv (\bV^T \bV)^{-1} \bV^T \bZ 
        &= \bv (\bV^T \bV)^{-1} \bV^T (\bAone + \bAtwo + \bAthree + \bB)\\
        &= \sum_{i=1}^n \lambda_i (A_{i1}+A_{i2}+A_{i3}+B_{i})
    \end{align*}
    where $\lambda_i$ denotes the $i$-th element of $\bv(\bV^T \bV)^{-1} \bV^T$. Note that $(\bV^T\bV)^{-1}$ is the kernel of the variance-covariance matrix of the OLS estimator which is the order of $1/n$ in general. Since $(\bV^T\bV)^{-1}$ is a $q\times q$ matrix, each element of the matrix and $\lambda_i$ are also the order of $1/n$. We write
    \begin{align*}
        &\sup\limits_{P_0} E_{P_0} \mathbb{P}_n\left(|\Delta(\boldsymbol{D}, \boldsymbol{\Psi})-\tau(\bv)|>M\epsilon_n|\bD\right)\\
        &=\sup\limits_{P_0} E_{P_0} \mathbb{P}_n\left(\left|\sum_{i=1}^n \lambda_i (A_{i1}+A_{i2}+A_{i3}+B_{i})-\tau(\bv)\right|>M\epsilon_n|\bD\right)\\
        &\leq\sup\limits_{P_0} E_{P_0} \mathbb{P}_n\left(\left|\sum_{i=1}^n \lambda_iA_{i1}\right|>M\epsilon_n/4|\bD\right)
        +\sup\limits_{P_0} E_{P_0} \mathbb{P}_n\left(\left|\sum_{i=1}^n \lambda_iA_{i2}\right|>M\epsilon_n/4|\bD\right)\\
        &\qquad+\sup\limits_{P_0} E_{P_0} \mathbb{P}_n\left(\left|\sum_{i=1}^n\lambda_iA_{i3}\right|>M\epsilon_n/4|\bD\right)
        +\sup\limits_{P_0} E_{P_0} \mathbb{P}_n\left(\left|\sum_{i=1}^n \lambda_iB_{i}-\tau(\bv)\right|>M\epsilon_n/4|\bD\right).
    \end{align*}
    Thus, it suffices to show under which conditions each of the four components above contracts at the $\epsilon_n$ rate. 
    
    We begin with the $B$ term. One can notice that it does not depend on the posterior distribution and all elements in the term are constant conditioning on $\bD$. Therefore, we  write
    \begin{align*}
        &\sup\limits_{P_0} E_{P_0} \mathbb{P}_n\left(\left|\sum_{i=1}^n \lambda_iB_{i}-\tau(\bv)\right|>M\epsilon_n/4|\bD\right)\\
        &=\sup\limits_{P_0} E_{P_0} 1\left(\left|\sum_{i=1}^n\lambda_iB_{i}-\tau(\bv)\right|>M\epsilon_n/4|\bD\right)\\
        &=\sup\limits_{P_0} P_{P_0} \left(\left|\sum_{i=1}^n\lambda_iB_i-\tau(\bv)\right|>M\epsilon_n/4|\bD\right)
    \end{align*}
    It can be shown that the quantity in the absolute value have expectation 0 with respect to $P_0$ if either one of the nuisance models is correctly specified, i.e. $\tilde{p}_{ti}=p_{ti}^*$ or $\tilde{m}_{ti}=m_{ti}^*$. Note first that if $\tilde{p}_{ti}=p_{ti}^*$, we can write  
    \begin{align*}
        \sum_{i=1}^nE_{P_0}\left[\lambda_i\left(\frac{1(T_i = t)}{\tilde{p}_{ti}} (Y_i - \tilde{m}_{ti}) + \tilde{m}_{ti}\right)\right]
        &=\sum_{i=1}^nE_{P_0}\left[\lambda_i\left(\frac{1(T_i = t)}{p_{ti}^*} (Y_i - \tilde{m}_{ti}) + \tilde{m}_{ti}\right)\right]\\
        &=\sum_{i=1}^nE_{X}E_{T|X}E_{Y|T,X}\left[\lambda_i\left(\frac{1(T_i = t)}{p_{ti}^*} (Y_i - \tilde{m}_{ti}) + \tilde{m}_{ti}\right)\right]\\
        &=\sum_{i=1}^nE_{X}E_{T|X}\left[\lambda_i\left(\frac{1(T_i = t)}{p_{ti}^*} (m_{ti}^* - \tilde{m}_{ti}) + \tilde{m}_{ti}\right)\right]\\
        &=\sum_{i=1}^nE_{X}\left[\lambda_i\left(\frac{p_{ti}^*}{p_{ti}^*} (m_{ti}^* - \tilde{m}_{ti}) + \tilde{m}_{ti}\right)\right]\\
        &=\sum_{i=1}^nE_{X}[\lambda_i m_{ti}^*]
    \end{align*}
    and similarly, if $\tilde{m}_{ti}=m_{ti}^*$,
    \begin{align*}
        \sum_{i=1}^nE_{P_0}\left[\lambda_i\left(\frac{1(T_i = t)}{\tilde{p}_{ti}} (Y_i - \tilde{m}_{ti}) + \tilde{m}_{ti}\right)\right]
        &=\sum_{i=1}^nE_{P_0}\left[\lambda_i\left(\frac{1(T_i = t)}{\tilde{p}_{ti}} (Y_i - m_{ti}^*) + m_{ti}^*\right)\right]\\
        &=\sum_{i=1}^nE_{X}E_{T|X}E_{Y|T,X}\left[\lambda_i\left(\frac{1(T_i = t)}{\tilde{p}_{ti}} (Y_i - m_{ti}^*) + m_{ti}^*\right)\right]\\
        &=\sum_{i=1}^nE_{X}E_{T|X}\left[\lambda_i\left(\frac{1(T_i = t)}{\tilde{p}_{ti}} (m_{ti}^* - m_{ti}^*) + m_{ti}^*\right)\right]\\
        &=\sum_{i=1}^nE_{X}[\lambda_i m_{ti}^*].
    \end{align*}
    Therefore, if either $\tilde{p}_{ti}=p_{ti}^*$ or $\tilde{m}_{ti}=m_{ti}^*$,
    \begin{align*}
        E_{P_0}\left[\sum_{i=1}^n\lambda_iB_i-\tau(\bv)\right]
        &=\sum_{i=1}^nE_{X}[\lambda_i(m_{1i}^*-m_{0i}^*)-\tau(\bv)]\\
        &=\sum_{i=1}^nE_{X}[\lambda_i\tau(\bv_i)-\tau(\bv)]\\
        &=E_{X} \Bigg[\sum_{i=1}^n \lambda_i\tau(\bv_i)-\tau(\bv) \Bigg]\\
        &=E_{X}[\tau(\bv)-\tau(\bv)]\\
        &=0
    \end{align*}
    The fourth equality follows from the fact that $\sum_{i=1}^n\lambda_i\tau(\bv_i)=\bv (\bV^T \bV)^{-1} \bV^T \boldsymbol{\tau}=\bv (\bV^T \bV)^{-1} \bV^T \bV \boldsymbol{\beta}=\bv\boldsymbol{\beta}=\tau(\bv)$ where $\boldsymbol{\tau}$ is a $n\times 1$ vector of $\tau(\bv_i)=\bv_i\boldsymbol{\beta}$. Therefore, we can apply the Chebyshev's inequality:
    \begin{align*}
        \sup\limits_{P_0} E_{P_0} \mathbb{P}_n\left(\left|\sum_{i=1}^n \lambda_iB_{i}-\tau(\bv)\right|>M\epsilon_n/4|\bD\right)
        &\leq\sup\limits_{P_0}\dfrac{16}{M^2\epsilon_n^2}\text{Var}_{P_0}\left(\sum_{i=1}^n \lambda_i B_{i}-\tau(\bv)\right)\\
        &\leq\sup\limits_{P_0}\dfrac{16}{M^2\epsilon_n^2}\dfrac{K_B}{n}
    \end{align*}

    which goes to zero if $\epsilon_n<n^{-1/2}$. We now show that $\text{Var}_{P_0}\left(\sum_{i=1}^n \lambda_i B_{i}-\tau(\bv)\right) \leq K_B/n$ for some $K_B$. First, we can see that we can write
    \begin{align*}
        \sum_{i=1}^n \lambda_i B_{i} -\tau(\bv) &= \bv(\bV^T\bV)^{-1}\bV^T\bB-\bv\boldsymbol{\beta}\\
        &=\bv[(\bV^T\bV)^{-1}\bV^T\bB-\boldsymbol{\beta}].
    \end{align*}
    Therefore, we can write
    \begin{align*}
        \text{Var}_{P_0}\left(\sum_{i=1}^n \lambda_i B_{i}-\tau(\bv)\right)
        &=\text{Var}_{P_0}\left(\bv[(\bV^T\bV)^{-1}\bV^T\bB-\boldsymbol{\beta}]\right)\\
        &=\bv\text{Var}_{P_0}\left((\bV^T\bV)^{-1}\bV^T\bB\right)\bv^T.
    \end{align*}
    The second equality holds because $\bv$ and $\boldsymbol{\beta}$ does not vary over the data generating distribution. We observe that $(\bV^T\bV)^{-1}\bV^T\bB$ is a least square estimate and therefore each element of its variance-covariance matrix is of order $1/n$. Since $\bv$ is a $q$-dimensional, it follows that $\bv\text{Var}_{P_0}\left((\bV^T\bV)^{-1}\bV^T\bB\right)\bv^T$ is also the order of $1/n$ as desired. Thus, we conclude the $\bB$ term contracts at the rate of $n^{-1/2}$ when either the outcome model or the propensity score model is correctly specified.
    
    Next, $A_1$ and $A_2$ terms have analogous proofs and thus we will only show the proof for $A_1$ for brevity. Throughout the rest of the proof, we focus on one side of pseudo-outcomes $Z^{(t)}$ and it is straightforward to extend to $Z=Z^{(1)}-Z^{(0)}$. First, we consider the case where the propensity score is misspecified. Then, we write
    \begin{align*}
        &\sup\limits_{P_0} E_{P_0} \mathbb{P}_n\left(\left|\sum_{i=1}^n \lambda_iA_{1i}^{(t)}\right|>M\epsilon_n/4|\bD\right)\\
        &=\sup\limits_{P_0} E_{P_0} \mathbb{P}_n\left(\left|\sum_{i=1}^n\lambda_i (m_{ti} - \tilde{m}_{ti}) \Bigg(1 - \frac{1(T_i = t)}{\tilde{p}_{ti}} \Bigg)\right|>M\epsilon_n/4|\bD\right)\\
        &=\sup\limits_{P_0} E_{P_0} \mathbb{P}_n\left(\left|\sum_{i=1}^n\dfrac{1}{n}\lambda'_i (m_{ti} - \tilde{m}_{ti}) \Bigg(1 - \frac{1(T_i = t)}{\tilde{p}_{ti}} \Bigg)\right|>M\epsilon_n/4|\bD\right)\\
        &\leq\sup\limits_{P_0} E_{P_0} \mathbb{P}_n\left(\sqrt{\dfrac{1}{n}\sum_{i=1}^n (\lambda'_i)^2(m_{ti} - \tilde{m}_{ti})^2} \sqrt{\dfrac{1}{n}\sum_{i=1}^n\Bigg(1 - \frac{1(T_i = t)}{\tilde{p}_{ti}} \Bigg)^2}>M\epsilon_n/4|\bD\right)\\
        &\leq\sup\limits_{P_0} E_{P_0} \mathbb{P}_n\left(\sqrt{\dfrac{1}{n}\sum_{i=1}^n (m_{ti} - \tilde{m}_{ti})^2} >\dfrac{M\epsilon_n}{4K}|\bD\right)\\
        &=\sup\limits_{P_0} E_{P_0} \mathbb{P}_n\left(\dfrac{1}{\sqrt{n}}\Vert \boldsymbol{m}_{t} - \boldsymbol{\tilde{m}}_{t}\Vert >\dfrac{M\epsilon_n}{4K}|\bD\right)
    \end{align*}
    where $\lambda'_i\equiv n\lambda_i$. The first inequality follows from the Cauchy-Schwartz inequality and the second inequality comes from the positivity assumption and the fact that $\lambda'_i$ is the order of 1. We can notice that the last term is just the posterior contraction rate of the outcome model. Therefore, if the propensity score is misspecified, the  $A_1$ term contracts at the rate that the outcome model contracts at regardless of correct specification of the outcome model. 
    
    Now let's consider the case where the propensity score is correctly specified. We can still write the quantity of interest as
    
       \begin{align*}
        &\sup\limits_{P_0} E_{P_0} \mathbb{P}_n\left(\left|\sum_{i=1}^n \lambda_iA_{1i}^{(t)}\right|>M\epsilon_n/4|\bD\right)\\
        &=\sup\limits_{P_0} E_{P_0} \mathbb{P}_n\left(\left|\sum_{i=1}^n\lambda_i (m_{ti} - \tilde{m}_{ti}) \Bigg(1 - \frac{1(T_i = t)}{\tilde{p}_{ti}} \Bigg)\right|>M\epsilon_n/4|\bD\right).
    \end{align*}
    
    Now we will use data splitting and the fact that the samples used to estimate the nuisance parameters are independent of the samples used for estimating the treatment effect. For now, we assume a fixed value of the nuisance parameters for $m_{ti}$, which means that we can define $l(\boldsymbol{X}_i) = m_{ti} - \tilde{m}_{ti}$, where $l(\cdot)$ is a function of the fixed nuisance parameters and the covariates for the samples used for estimating the causal effect. We can now write this expression as
    \begin{align*}
        &\sup\limits_{P_0} E_{P_0} \mathbb{P}_n\left(\left|\sum_{i=1}^n\lambda_i (m_{ti} - \tilde{m}_{ti}) \Bigg(1 - \frac{1(T_i = t)}{\tilde{p}_{ti}} \Bigg)\right|>M\epsilon_n/4|\bD\right) \\
        &= \sup\limits_{P_0} E_{P_0} \mathbb{P}_n\left(\left|\sum_{i=1}^n\lambda_i l(\boldsymbol{X}_i) \Bigg(1 - \frac{1(T_i = t)}{\tilde{p}_{ti}} \Bigg)\right|>M\epsilon_n/4|\bD\right) \\
        &=\sup\limits_{P_0} E_{P_0} 1\left(\left|\sum_{i=1}^n\lambda_i l(\boldsymbol{X}_i) \Bigg(1 - \frac{1(T_i = t)}{\tilde{p}_{ti}} \Bigg)\right|>M\epsilon_n/4|\bD\right) \\
        &=\sup\limits_{P_0} P_{P_0} \left(\left|\sum_{i=1}^n\lambda_i l(\boldsymbol{X}_i) \Bigg(1 - \frac{1(T_i = t)}{\tilde{p}_{ti}} \Bigg)\right|>M\epsilon_n/4|\bD\right).
    \end{align*}
    
    Now we can show that the quantity inside of the absolute value has expectation zero with respect to $P_0$ when the propensity score is correctly specified:
    \begin{align*}
        & E_{P_0} \Bigg(\sum_{i=1}^n\lambda_i l(\boldsymbol{X}_i) \bigg(1 - \frac{1(T_i = t)}{\tilde{p}_{ti}} \bigg) \Bigg) \\
        &= \sum_{i=1}^n E_{P_0} \Bigg( \lambda_i l(\boldsymbol{X}_i) \bigg(1 - \frac{1(T_i = t)}{\tilde{p}_{ti}} \bigg) \Bigg) \\
        &= \sum_{i=1}^n E_X E_{T|X} E_{Y | T,X} \Bigg( \lambda_i l(\boldsymbol{X}_i) \bigg(1 - \frac{1(T_i = t)}{\tilde{p}_{ti}} \bigg) \Bigg) \\
        &= \sum_{i=1}^n E_X E_{T|X}  \Bigg( \lambda_i l(\boldsymbol{X}_i) \bigg(1 - \frac{1(T_i = t)}{\tilde{p}_{ti}} \bigg) \Bigg) \\
        &= \sum_{i=1}^n E_X  \Bigg( \lambda_i l(\boldsymbol{X}_i) \bigg(1 - \frac{p_{ti}^*}{\tilde{p}_{ti}} \bigg) \Bigg) \\
        &= 0,
    \end{align*}
    where the last equality followed because $p_{ti}^* = \tilde{p}_{ti}$ when the propensity score is correctly specified. Since this quantity has mean zero, we can apply Chebyshev's inequality to say that 
    \begin{align*}
        &\sup\limits_{P_0} P_{P_0} \left(\left|\sum_{i=1}^n\lambda_i l(\boldsymbol{X}_i) \Bigg(1 - \frac{1(T_i = t)}{\tilde{p}_{ti}} \Bigg)\right|>M\epsilon_n/4|\bD\right) \\
        & \leq \sup\limits_{P_0}\dfrac{16}{M^2\epsilon_n^2}\text{Var}_{P_0}\left(\sum_{i=1}^n\lambda_i l(\boldsymbol{X}_i) \Bigg(1 - \frac{1(T_i = t)}{\tilde{p}_{ti}} \Bigg)\right)\\
        &= \sup\limits_{P_0}\dfrac{16}{M^2\epsilon_n^2}\text{Var}_{P_0}\left(\frac{1}{n} \sum_{i=1}^n\lambda_i' l(\boldsymbol{X}_i) \Bigg(1 - \frac{1(T_i = t)}{\tilde{p}_{ti}} \Bigg)\right)\\
        &= \sup\limits_{P_0}\dfrac{16}{M^2\epsilon_n^2 n^2}\text{Var}_{P_0}\left( \sum_{i=1}^n\lambda_i' l(\boldsymbol{X}_i) \Bigg(1 - \frac{1(T_i = t)}{\tilde{p}_{ti}} \Bigg)\right)\\
        & \leq \sup\limits_{P_0}\dfrac{16 K_{A_1}}{ M^2\epsilon_n^2 n}.
    \end{align*}
    One can easily justify the last inequality using a similar idea to the $\bB$ term. We can now see that this quantity approaches zero as long as $\epsilon_n > n^{-1/2}$, which shows that this term contracts to zero at the $n^{-1/2}$ rate when the propensity score is correctly specified. Note that this relied on treating $m_{ti} - \tilde{m}_{ti}$ as a fixed quantity that we referred to as $l(\boldsymbol{X}_i)$. This term is random, however, and can vary according to the posterior distribution for the outcome regression model. This result only required that $l(\boldsymbol{X}_i)$ was bounded, which ensured that the variance above is finite. According to our assumption on the posterior distribution for the outcome model, this is guaranteed to be finite and therefore the result holds for any value of the nuisance parameters with nonzero density under the posterior distribution for $m_{ti}$, thereby ensuring the result holds in general. Note that we used data splitting in this proof to ensure that the data used to estimate the nuisance parameters was independent of the data used to estimate the causal effect, which in turn let us write $m_{ti} - \tilde{m}_{ti}$ as $l(\boldsymbol{X}_i)$. This also keeps us from needing to make any assumptions (other than consistency) on the propensity score or outcome regression models. Alternatively, this result could be proven assuming these models fall within a Donsker class of functions (see Lemma 19.24 of \cite{Vaart1998}), though we do not explicitly show that here.

    We move to the $A_3$ term which involves both posterior distributions. We start with writing
    \begin{align*}
        &\sup\limits_{P_0} E_{P_0} \mathbb{P}_n\left(\left|\sum_{i=1}^n \lambda_iA_{i3}^{(t)}\right|>M'\epsilon_n/4|\bD\right)\\
        &=\sup\limits_{P_0} E_{P_0} \mathbb{P}_n \left( \left|\sum_{i=1}^n \lambda_i\frac{1(T_i = t)(p_{ti} - \tilde{p}_{ti}) (m_{ti} - \tilde{m}_{ti})}{p_{ti}\tilde{p}_{ti}}\right| > M'\epsilon_n/4 \vert \boldsymbol{D} \right) \\
        &=\sup\limits_{P_0} E_{P_0} \mathbb{P}_n \left( \left|\sum_{i=1}^n \dfrac{1}{n}\lambda'_i\frac{1(T_i = t)(p_{ti} - \tilde{p}_{ti}) (m_{ti} - \tilde{m}_{ti})}{p_{ti}\tilde{p}_{ti}}\right| > M'\epsilon_n/4 \vert \boldsymbol{D} \right) \\
        &\leq\sup\limits_{P_0} E_{P_0} \mathbb{P}_n \left( \dfrac{1}{n}\sqrt{\sum_{i=1}^n (\lambda'_i)^2\left( \frac{1(T_i = t)(p_{ti} - \tilde{p}_{ti}) }{p_{ti}\tilde{p}_{ti}}\right)^2}\sqrt{\sum_{i=1}^n (m_{ti} - \tilde{m}_{ti})^2}  >M'\epsilon_n/4 \vert \boldsymbol{D} \right)\\
        &\leq\sup\limits_{P_0} E_{P_0} \mathbb{P}_n \left( \dfrac{1}{n}\sqrt{\sum_{i=1}^n K_{3}(p_{ti} - \tilde{p}_{ti})^2}\sqrt{\sum_{i=1}^n(m_{ti} - \tilde{m}_{ti})^2} > M' \epsilon_{n}/4 \vert \boldsymbol{D} \right)\\
        &=\sup\limits_{P_0} E_{P_0} \mathbb{P}_n \left( \dfrac{1}{n}\Vert \boldsymbol{p}_t-\boldsymbol{\tilde{p}}_t\Vert \Vert \boldsymbol{m}_t-\boldsymbol{\tilde{m}}_t\Vert > \dfrac{M' \epsilon_{n}}{4\sqrt{K_3}} \vert \boldsymbol{D} \right)
    \end{align*}
    where the first inequality holds by Cauchy-Schwartz inequality and the second inequality is true for some constant $0<K_3<\infty$ by the positivity and the bounded error assumptions combining with the fact that $\lambda'_i$ is the order of $1$. We further decompose this probability as
    \begin{align*}
        &\sup\limits_{P_0} E_{P_0} \Bigg[\mathbb{P}_n \left( \dfrac{1}{n}\Vert \boldsymbol{p}_t-\boldsymbol{\tilde{p}}_t\Vert \Vert \boldsymbol{m}_t-\boldsymbol{\tilde{m}}_t\Vert > \dfrac{M' \epsilon_{n}}{4\sqrt{K_3}} \vert \boldsymbol{D}, \dfrac{1}{\sqrt{n}}\Vert \boldsymbol{m}_t-\boldsymbol{\tilde{m}}_t\Vert> \epsilon_n^{\nu} \right)\\
        &\hspace{1 in}\times\mathbb{P}_n \left(\dfrac{1}{\sqrt{n}}\Vert \boldsymbol{m}_t-\boldsymbol{\tilde{m}}_t\Vert> \epsilon_n^{\nu} \right)\\
        &\hspace{.7 in}+\mathbb{P}_n \left( \dfrac{1}{n}\Vert \boldsymbol{p}_t-\boldsymbol{\tilde{p}}_t\Vert \Vert \boldsymbol{m}_t-\boldsymbol{\tilde{m}}_t\Vert > \dfrac{M' \epsilon_{n}}{4\sqrt{K_3}} \vert \boldsymbol{D}, \dfrac{1}{\sqrt{n}}\Vert \boldsymbol{m}_t-\boldsymbol{\tilde{m}}_t\Vert \leq \epsilon_n^{\nu} \right)\\
        &\hspace{1 in}\times\mathbb{P}_n \left(\dfrac{1}{\sqrt{n}}\Vert \boldsymbol{m}_t-\boldsymbol{\tilde{m}}_t\Vert \leq \epsilon_n^{\nu} \right)\Bigg]\\
        &\leq\sup\limits_{P_0} E_{P_0} \Bigg[\mathbb{P}_n \left(\dfrac{1}{\sqrt{n}}\Vert \boldsymbol{m}_t-\boldsymbol{\tilde{m}}_t\Vert> \epsilon_n^{\nu} \right)\\
        &\hspace{.7 in}+\mathbb{P}_n \left( \dfrac{1}{n}\Vert \boldsymbol{p}_t-\boldsymbol{\tilde{p}}_t\Vert \Vert \boldsymbol{m}_t-\boldsymbol{\tilde{m}}_t\Vert > \dfrac{M' \epsilon_{n}}{4\sqrt{K_3}} \vert \boldsymbol{D}, \dfrac{1}{\sqrt{n}}\Vert \boldsymbol{m}_t-\boldsymbol{\tilde{m}}_t\Vert \leq \epsilon_n^{\nu} \right)\Bigg]\\
        &\leq\sup\limits_{P_0} E_{P_0} \Bigg[\mathbb{P}_n \left(\dfrac{1}{\sqrt{n}}\Vert \boldsymbol{m}_t-\boldsymbol{\tilde{m}}_t\Vert> \epsilon_n^{\nu} \right)+\mathbb{P}_n \left( \dfrac{1}{\sqrt{n}}\Vert \boldsymbol{p}_t-\boldsymbol{\tilde{p}}_t\Vert > \dfrac{M' \epsilon_{n}^{1-\nu}}{4\sqrt{K}} \vert \boldsymbol{D} \right)\Bigg]
    \end{align*}
    The first term converges to zero if the outcome model contracts at the $\epsilon_n^{\nu}$ rate, while the second term converges to zero if the propensity score model converges at the $1-\epsilon_n^{\nu}$ rate. Overall, this shows that the $A_3$ term contracts at the $\epsilon_n$ rate as long as the product of contraction rates of two nuisance models is $\epsilon_n$ regardless of correct specification of models.
    
    Now we look at the overall contraction rate. If both models are correctly specified, then the $A_1$, $A_2$ and $B$ terms contract at $\epsilon_n=n^{-1/2}$ rate, while $A_3$ contracts at $\epsilon_{nt}\epsilon_{ny}$, the product of two contraction rates of nuisance models. Thus, the doubly robust estimator contracts at either $n^{-1/2}$ or $\epsilon_{nt}\epsilon_{ny}$, whichever is slower. If either one of the nuisance models is misspecified, then, depending on which model is misspecified, one of the $A_1$ and $A_2$ terms contracts at $n^{-1/2}$ and the other contracts at the contraction rate of the correctly specified model ($\epsilon_{nt}$ or $\epsilon_{ny}$) while the contraction rates of $A_3$ and $B$ terms do not change. Thus, the overall contraction rate will be either $n^{-1/2}$ or the contraction rate of the correctly specified model ($\epsilon_{nt}$ or $\epsilon_{ny}$), whichever is slower.
    
    In summary, if either one of the nuisance models is correctly specified, the posterior distribution of $\Delta(\boldsymbol{D}, \boldsymbol{\Psi})$ contracts to the true treatment effect $\tau(\bv)$. Therefore, the posterior median of $\Delta(\boldsymbol{D}, \boldsymbol{\Psi})$ is consistent for $\tau(\bv)$. Further, assuming that the posterior variance of $\Delta(\boldsymbol{D}, \boldsymbol{\Psi})$ is bounded, which is mild, we can also conclude that our estimator $E_{\bPsi|\bD}[\Delta(\boldsymbol{D}, \boldsymbol{\Psi})]$ is consistent for $\tau(\bv)$.

\section{Normal approximation of the doubly robust estimator}
\label{sec:normal approximation}

We borrow the decomposition of $Z_i$ from Appendix \ref{sec:consistency} and focus on the difference between two pseudo-outcomes. To find the limiting distribution of our estimator, we first take a look at $\bB$. Since $\bB$ does not depend on the posterior distribution, we write $E_{\bPsi | \bD} \left\{ \bv (\bV^T \bV)^{-1} \bV^T \bB \right\}= \bv (\bV^T \bV)^{-1} \bV^T \bB$. Observe that $(\bV^T \bV)^{-1} \bV^T \bB$ is the regression coefficient of the linear regression model of $\bB$ onto $\bV$. Recall that $m_{1i}^*-m_{0i}^*=\tau(\bv_i)=\bv_i\boldsymbol{\beta}$ assuming linear CATE. Then, we can re-write
    \begin{align*}
        B_i&=\begin{cases}
        (m_{1i}^*-m_{0i}^*)+\frac{1}{p_{1i}^*} (Y_i - m_{1i}^*)\qquad \text{if $T_i=1$}\\
        (m_{1i}^*-m_{0i}^*)-\frac{1}{p_{0i}^*} (Y_i - m_{0i}^*)\qquad \text{if $T_i=0$,}
        \end{cases}\\
        &=\bv_i\boldsymbol{\beta}+\epsilon_i
    \end{align*}
    where $\epsilon_i$ is $i.i.d$ and takes a value either $\frac{1}{p_{1i}^*} (Y_i - m_{1i}^*)$ or $-\frac{1}{p_{0i}^*} (Y_i - m_{0i}^*)$. Further, recall that  $E(Y_i|T_i=t,\bX_i)=m_{ti}^*$ and $0<p_{ti}<1$. Therefore, $E\left(\epsilon_i|T_i=t,\bX_i\right)=0$ and the variance of $\epsilon_i$ is finite.
    It is well-known that the regression coefficient of the linear regression model is asymptotically normal if 1) $\epsilon_i$'s are $i.i.d$ with mean $0$ and a finite variance, 2) the matrix $\bV$ is uniformly bounded, 3) and $\lim_{n\rightarrow\infty} \bV^T\bV/n$ is finite and non-singular. We have already seen that the first condition holds in our case. The other conditions are quite weak and already assumed in the manuscript. 
    Therefore, we can conclude that $\bv (\bV^T \bV)^{-1} \bV^T \bB$ is asymptotically normally distributed.
    
    Now, we look at the other terms. As we proved in Appendix \ref{sec:behavior of the true variance}, under correct specification of both nuisance models,
    \begin{align*}
        \text{Var}_{\boldsymbol{D}} \{ E_{\boldsymbol{\Psi} \vert \boldsymbol{D}} [\bv (\bV^T \bV)^{-1} \bV^T \boldsymbol{A}_1] \} &= \text{Var}_{\boldsymbol{D}} \{ E_{\boldsymbol{\Psi} \vert \boldsymbol{D}} [\bv (\bV^T \bV)^{-1} \bV^T \boldsymbol{A}_2] \} \\
        &= \text{Var}_{\boldsymbol{D}} \{ E_{\boldsymbol{\Psi} \vert \boldsymbol{D}} [\bv (\bV^T \bV)^{-1} \bV^T \boldsymbol{A}_3] \}\\
        &=o_p(n^{-1}).
    \end{align*}
    In other words, the variances of the terms related to $\boldsymbol{A}_1,\boldsymbol{A}_2$, and $\boldsymbol{A}_3$ are asymptotically negligible. This, combined with the consistency results of Appendix \ref{sec:consistency of estimator}, leads us to conclude that $$E_{\bPsi | \bD} \Big[ \Delta(\boldsymbol{D}, \boldsymbol{\Psi}) \Big]\overset{d}{\rightarrow}N(\tau(\bv),\text{Var}_{\boldsymbol{D}}[\bv (\bV^T \bV)^{-1} \bV^T \bB]).$$
    This holds under the same conditions as the proof in Appendix \ref{sec:consistency}, which are that both of the propensity score and outcome regression posterior distributions contract sufficiently fast towards the true parameters. 

\section{Comparison of Simulation Performances With and Without Sample Splitting}
\label{sec:with and without sample splitting}
\begin{figure}[ht]
    \centering
    \includegraphics{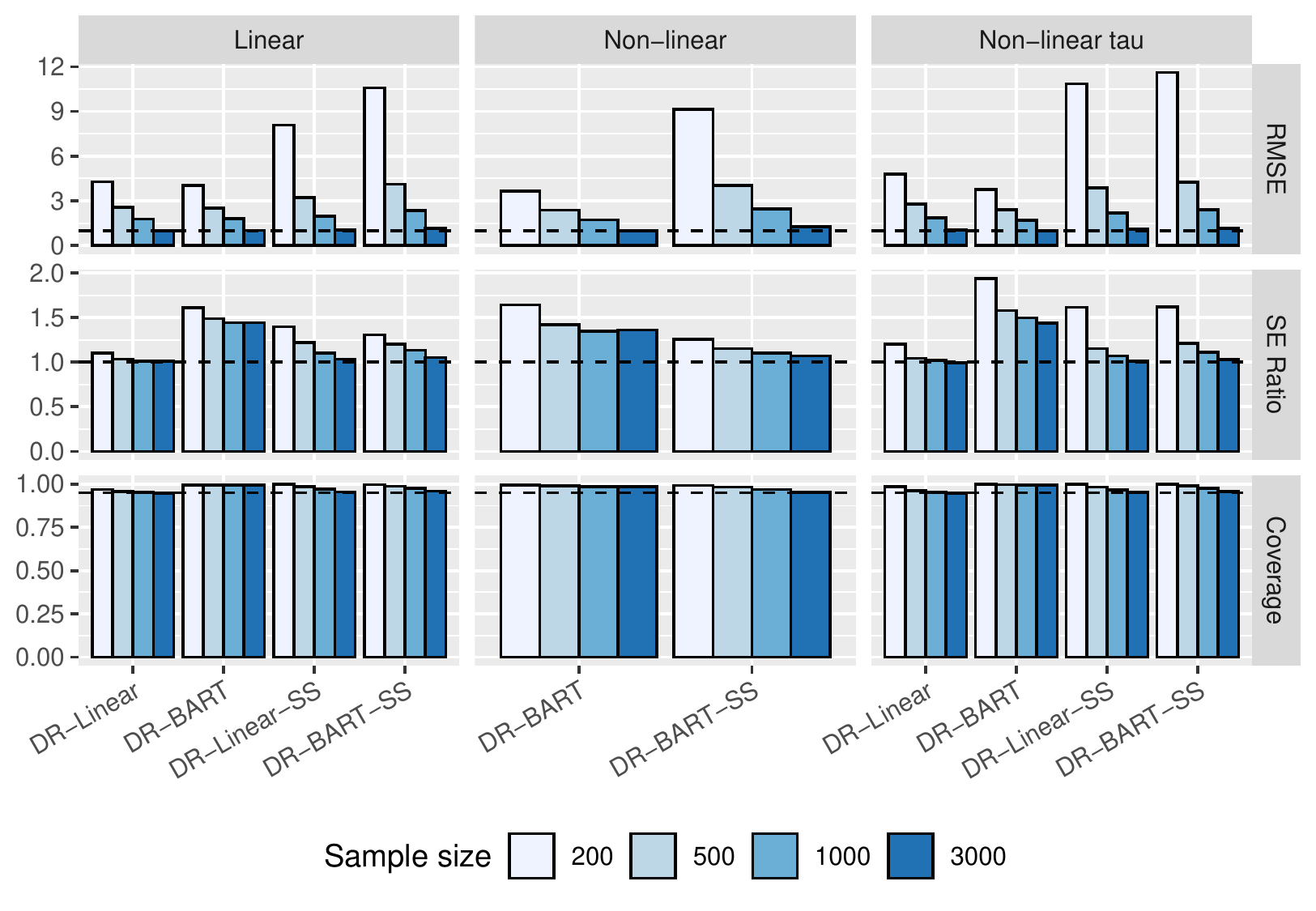}
    \caption{Comparison of the quality of the estimators with sample splitting (\texttt{DR-Linear-SS} and \texttt{DR-BART-SS}) and without it (\texttt{DR-Linear} and \texttt{DR-BART}). The first two columns correspond to Section \ref{sec:linear} and the third column corresponds to Section \ref{sec:Non-linear tau case}.}
    \label{fig:SampleSplit}
\end{figure}

\begin{figure}
    \centering
    \includegraphics{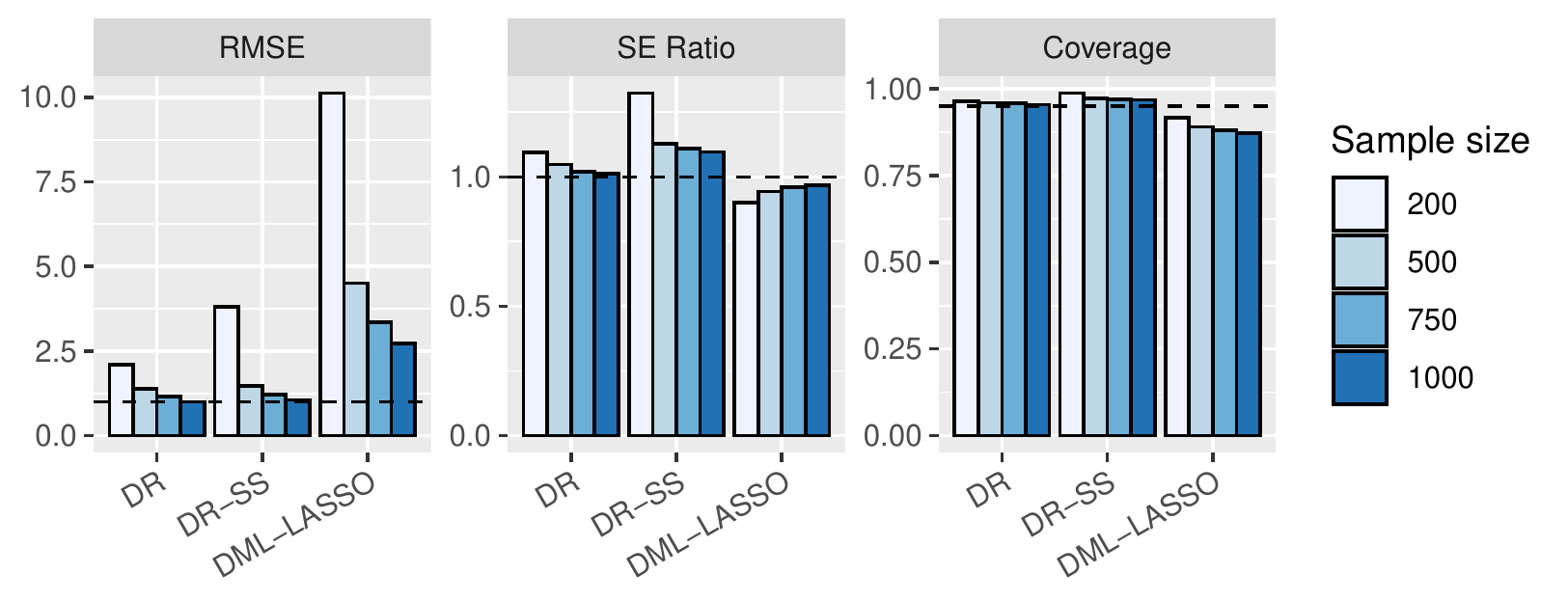}
    \caption{Comparison of the quality of the estimators with sample splitting (\texttt{DR-SS}) and without it (\texttt{DR}) along with the competing method (\texttt{DML-LASSO}) in high-dimensions in Section \ref{sec:High-dimensional nuisance functions}.}
    \label{fig:SS in HD}
\end{figure}

In this section, we examine how the estimators with and without sample splitting perform in simulations. For sample splitting estimators we split the data into two parts, where one is used to estimate the nuisance parameters and the other is used to estimate the CATE.  We can do this twice and reverse the order of which data split is used for parameter estimation and for CATE estimation. This gives us two estimators, which we can average to produce a single estimate of the CATE. For variance estimation, we can take the sum of the two separate variance estimates and divide by four. This should reduce the negative impact of splitting the data at each stage and reduce the RMSE of our estimator without sacrificing inferential properties. Figure \ref{fig:SampleSplit} and Figure \ref{fig:SS in HD} suggest that our theoretical results hold for our estimators both with and without sample splitting. We see conservative inference regardless of whether sample splitting is used, and a consistent estimator for the variance in situations where the posterior distributions contract sufficiently fast (\texttt{DR-Linear} and \texttt{DR}). In terms of RMSE, however, we find that the full data estimator that does not perform sample splitting is much better, particularly in terms of small sample performance, while our method with sample splitting performs better than a competing method in high-dimensions. As the sample size increases, the RMSE of the two approaches using and not using sample splitting are much more similar.


\section{Simulation Results for the Observed Data Locations}\label{sec: training}
We also can evaluate the CATE at the observed data locations, given by $\tau(\bV_i)$ for $i=1, \dots, n$. This allows us to compare to the Bayesian causal forest (\texttt{BCF}) approach, and we use the same data generating process and metrics as Section \ref{sec:linear}.
\begin{figure}[ht]
    \centering
    \includegraphics{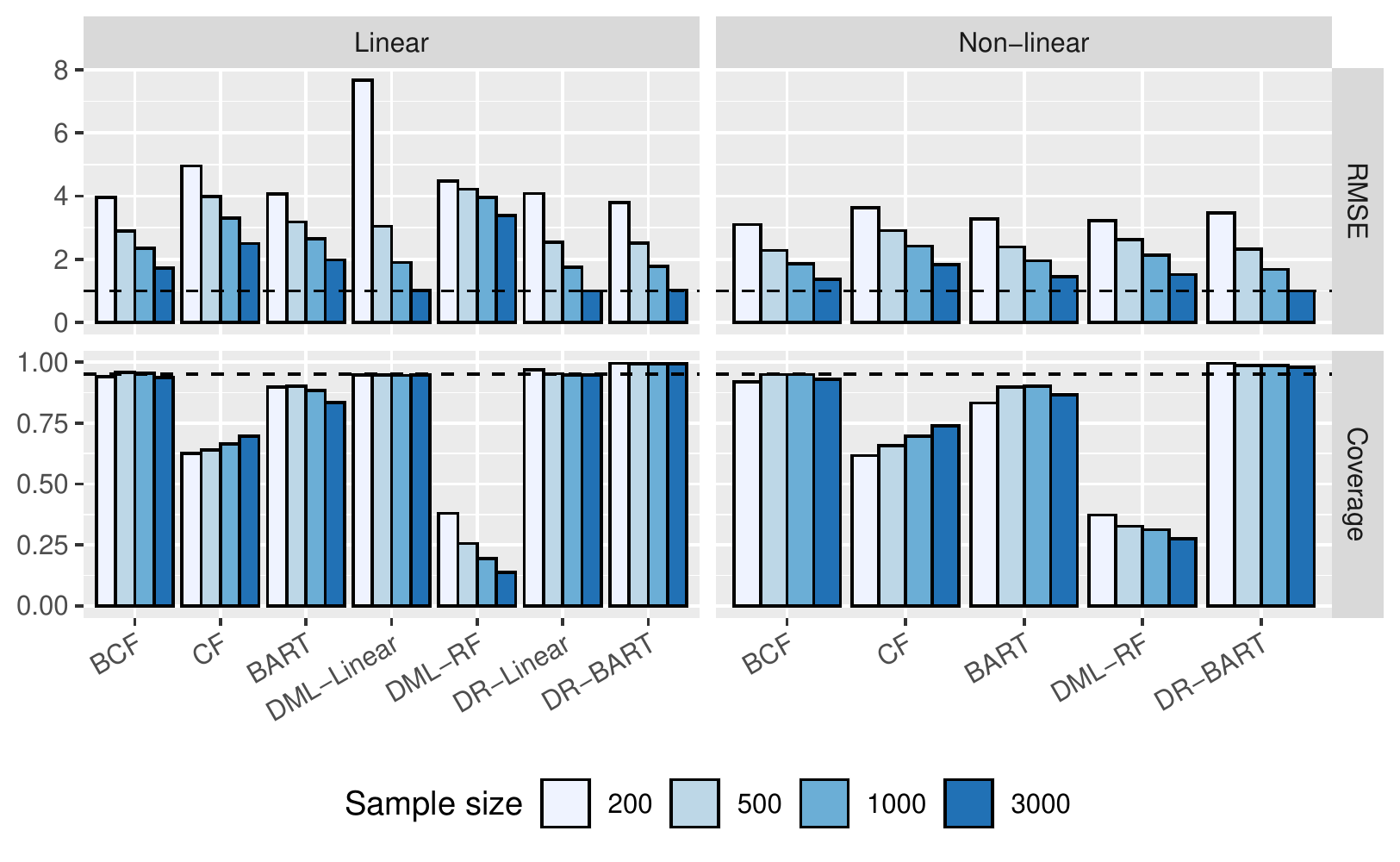}
    \caption{Results for the simulation in Section \ref{sec:linear} at the observed locations.}
    \label{fig:low-training}
\end{figure}
In Figure \ref{fig:low-training}, we find similar results to those seen in the main manuscript. The proposed methods produce valid inference regardless of the sample size, and generally have the lowest RMSE of all methods considered. We see that \texttt{BCF} outperforms \texttt{CF}, particularly in terms of uncertainty quantification as the interval coverages are much closer to the nominal 95\% level, though slightly below it in certain situations. Additionally the RMSE of \texttt{BCF} is competitive to the other estimators considered, though tends to be larger than the proposed approaches when the sample sizes are larger. Lastly, we only considered the low-dimensional simulation scenario here, as the \texttt{BCF} approach is not applicable to high-dimensional covariate spaces. 

\section{Results for Using the Bootstrap for High-dimensional and Nonparametric Models}\label{sec: Bootstrap}

\begin{figure}
    \centering
    \includegraphics{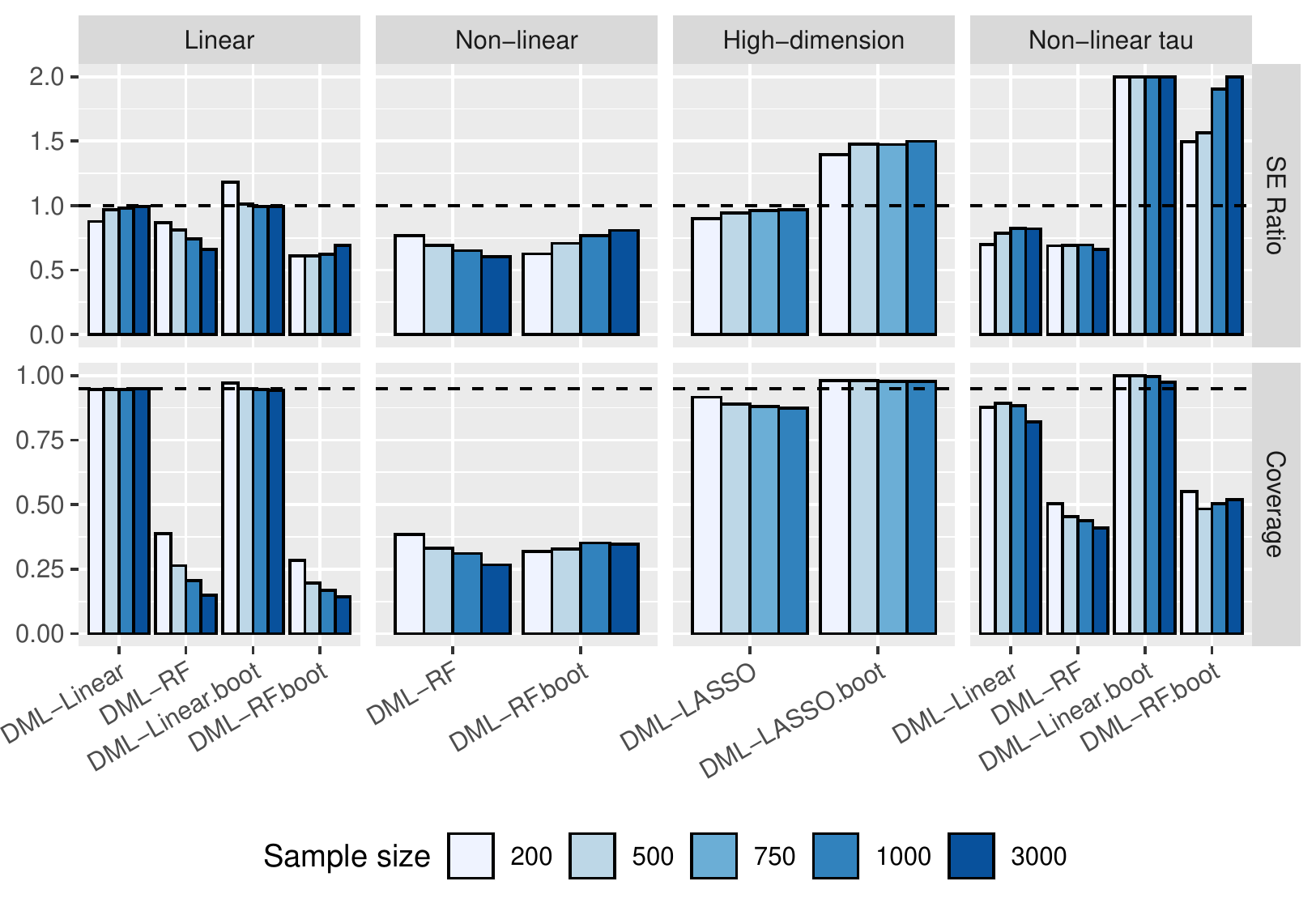}
    \caption{Comparing \texttt{DML} with and without bootstrapping. The first two columns correspond to Section \ref{sec:linear}, while the third and fourth columns correspond to Section \ref{sec:High-dimensional nuisance functions} and \ref{sec:Non-linear tau case}, respectively. The standard error ratios are truncated at 2 for readability.}
    \label{fig:Boot}
\end{figure}

Here we show the performance of the bootstrap for uncertainty quantification with the \texttt{DML} estimator. To do so, we compare the standard \texttt{DML} without the bootstrap (i.e. with the original variance estimation process) to the modified \texttt{DML} with the bootstrap denoted by \texttt{DML-*.boot}. As shown in Figure \ref{fig:Boot}, \texttt{DML} is improved by bootstrapping only when linear models are used for the nuisance parameters and the covariate space is low-dimensional. When either high-dimensional or nonparametric models such as random forests, LASSO, or spline are used, however, it fails to produce valid measures of uncertainty. Either the variance estimates are still underestimating the true uncertainty, or they are unreasonably large with standard error ratios that can far exceed 2. 

\end{document}